\documentclass[12pt]{article}

\usepackage{a4,a4wide}

\usepackage{amsmath}
\usepackage{amssymb} 
\usepackage{fancybox}
\usepackage{graphicx}
\usepackage[dvips]{color}
\makeatletter

\@addtoreset{equation}{section}
\makeatother

\usepackage[debug,colorlinks=true
,urlcolor=blue
,anchorcolor=blue
,citecolor=green
,filecolor=blue
,linkcolor=blue
,menucolor=blue
,pagecolor=blue
,linktocpage=true,pageanchor=false
]{hyperref}

\newtheorem{thm}{Theorem}[section]

\newtheorem{pf}[thm]{Proof}

\def\Tr{\mathrm{Tr}}

\def\half{{1\over2}}
\def\nn{\nonumber\\}

\def\[{\left[}
\def\]{\right]}
\def\({\left(}
\def\){\right)}

\newcommand{\ol}{\overline}

\def\={\stackrel{\bullet}{=}}
\def\nn{\notag\\}

\def\cL{{\cal L}}
\def\cN{{\cal N}}
\def\cO{{\cal O}}
\def\cP{{\cal P}}

\def\mbf{\mathbf}

\def\mf {\mathfrak }

\def \be {\begin{equation}}
\def \ee {\end{equation}}
\def \bea {\begin{eqnarray}}
\def \eea {\end{eqnarray}}
\def \beal#1 {\begin{align}#1\end{align}}
\def \bes#1 {\begin{equation}\begin{split}#1\end{split}\end{equation}}
\def \nn {\notag\\}
\def\sla#1{\not\!\!#1}

\def\ol#1{\overline{#1}}

\begin{document}

\begin{titlepage}
\vspace{-3cm}
\title{
\begin{flushright}
\normalsize{ 
June 2015}
\end{flushright}
       \vspace{1.5cm}
Supersymmetry Algebra in Super Yang-Mills Theories
       \vspace{1.5cm}
}
\author{
Shuichi Yokoyama
\\[25pt] 
\!\!\!\!\!\!\!\!{\it \normalsize Physics Department, Technion - Israel Institute of Technology,}\\
\!\!\!\!\!\!\!\!{\it \normalsize Technion City -
Haifa 3200003}\\
\\[10pt]
\!\!\!\!\!\!\!\!{\small \tt E-mail: shuichitoissho(at)gmail.com}
}

\date{}
\maketitle

\thispagestyle{empty}

\vspace{.2cm}

\begin{abstract}
\vspace{0.3cm}
\normalsize
We compute supersymmetry algebra (superalgebra) in supersymmetric Yang-Mills theories (SYM) consisting of a vector multiplet including fermionic contribution in six dimensions. 
We show that the contribution of fermion is given by boundary terms.
From six dimensional results we determine superalgebras of five and four dimensional SYM by dimensional reduction. 
In five dimensional superalgebra the Kaluza-Klein momentum and the instanton particle charge are not the same but algebraically indistinguishable. 
We also extend this calculation including a hyper multiplet and for maximally SYM.  
We derive extended supersymmetry algebras in these four dimensional SYM with the holomorphic coupling constant given in hep-th/9408099.

\end{abstract}
\end{titlepage}

\section{Introduction} 
\label{intro} 

Supersymmetric Yang-Mills theories (SYM) in higher dimensions than four \cite{Brink:1976bc} have been uncovered to possess their own rich structure of supersymmetric (SUSY) quantum field theories (QFT)
in spite of their nature of lack of power counting renormalizability. 

In five dimensional case the structure of Coulomb branch at long distance can be determined exactly due to the fact that prepotential can be computed exactly by one-loop \cite{Witten:1996qb,Seiberg:1996bd}. 
What was interestingly found is that if the number of matter multiplets is small enough, there is no singularity of Landau pole and it becomes possible to take strong coupling limit on smooth moduli space, which leads to an ultra-violet (UV) fixed point with global symmetry enhancement depending on the matter content. 
This phenomenon has been further studied by using brane construction \cite{Aharony:1997ju,Aharony:1997bh,DeWolfe:1999hj,Benini:2009gi,Bergman:2013aca}, a superconformal index \cite{Kim:2012gu,Bashkirov:2012re,Taki:2013vka,Bergman:2013ala,Bergman:2013koa} and direct state analysis \cite{Tachikawa:2015mha,Zafrir:2015uaa,Yonekura:2015ksa}.

Maximally SYM in five dimensions has also attracted a great deal of attention and studied in relation to six dimensional (2,0) superconformal field theory (SCFT) \cite{Douglas:2010iu,Lambert:2010iw}, whose Lagrangian description is unknown. 
Although it was shown that UV divergence of five dimensional SYM appears at six loops \cite{Bern:2012di}, which indicates necessity of UV completion, 
BPS sector of the theory is expected to encode information of that of (2,0) SCFT due to its insensitivity to UV. It was shown that five dimensional maximally SYM contains Kaluza-Klein modes coming from the sixth direction as states with instanton-particle charge \cite{Lambert:2010iw,Kim:2011mv,Kim:2012qf}.  

Search of a SUSY gauge theory enjoying a non-trivial UV fixed point has also been done in six dimensions \cite{Seiberg:1996qx}. 
The requirement is gauge anomaly cancellation as is the case in even dimensional QFT. 
It has been shown that anomaly of matter multiplets can cancel if the number is small enough for SU(2) gauge group. 
This was further studied in other simple gauge groups \cite{Danielsson:1997kt}. Examples of nontrivial UV fixed points are provided by compactification of string theory with strong coupling (or tensionless) limit \cite{Ganor:1996mu,Seiberg:1996vs,Duff:1996cf,Ganor:1996nf}. See \cite{Witten:1997kz} for other examples of six dimensional gauge theories. 

In comparison to these non-trivial developments of higher dimensional SUSY gauge theories this paper performs a basic calculation for an aim to determine supersymmetry algebra (superalgebra) of six dimensional SYM.   
Lagrangian description allows us to compute six dimensional superalgebra explicitly and dimensional reduction for the six dimensional result enables us to compare the Kaluza-Klein momentum of the sixth direction and instanton-particle charge, which are identified in earlier study. 
We also recover a basic result of superalgebra of four dimensional $\cN=2$ SYM including a hyper multiplet, which leads to the formula of central charge with the holomorphic coupling constant insightfully chosen in \cite{Seiberg:1994aj}.

The rest of this paper is organized as follows. 
In \S\ref{10d} we review the method to determine superalgebra by using ten dimensional SYM following \cite{Osborn:1979tq}. 
In \S\ref{6d} we compute superalgebra of SYM in six dimensions including contribution of a hyper mutliplet (\S\ref{6dHM}). In particular the $\cN=2$ algebra in six dimensions is determined by dimensional reduction of ten dimensional one.
In \S\ref{5d}, \S\ref{4d} we determine superalgebras of five and four dimensional SYM, respectively, by dimensional reduction from six or ten dimensions.  
\S\ref{discussion} is devoted to conclusion and discussion. 
Appendix contains a formula of gamma matrix (\S\ref{gamma}) and 
convention in six dimensions used in this paper (\S\ref{convention}). 

\section{Superalgebra in 10d SYM}
\label{10d}

In this section we review the supersymmetry algebra in ten dimensional supersymmetric Yang-Mills theory \cite{Osborn:1979tq} using our convention. 
Results in this section are used to derive similar results of maximally SYM in other dimensions by dimensional reduction  later.  
The fields of SYM in ten dimensions are 
a gauge field $ A_{\bf M}\, ({\bf M}=0, 1, \cdots, 9)$%
\footnote{ The gauge field in this paper is anti-hermitian. }
and a Majorana-Weyl fermion (gaugino) $\lambda$, whose chirality we choose as positive.
\be 
\hat\Gamma_{10} \lambda = \lambda, \quad 
\lambda = C_{10}\bar\lambda^T,
\label{10dMW}
\ee
where $\Gamma_\mbf M$ are SO(1,9) gamma matrices, $\bar\lambda=i\lambda^\dagger\Gamma_0$, 
\be 
\hat\Gamma_{10}=\Gamma_{01\cdots 9}, \quad 
C_{10}= -\Gamma_{03579}.
\ee 
We realize the ten dimensional gamma matrices by using six dimensional ones as \eqref{10dgamma} in \S\ref{6dn2}, which is useful for dimensional reduction carried out later.  
We employ matrix notation for spinor indices and $^T$ acts only on them. 
The SYM Lagrangian (density) in ten dimensions is given by 
\be
\cL_{10} ={1 \over g_{10}^2} \Tr\biggl[{1\over 4}  F_{{\bf M}{\bf N}} F^{{\bf M}{\bf N}} + \half \bar\lambda \Gamma^{\bf M} D_{\bf M} \lambda \biggr]
\label{10dLagrangian}
\ee
where $ F_{{\bf M}{\bf N}}= [D_{\bf M},D_{\bf N}]$, $D_{\bf M}= \partial_{\bf M}+ A_{\bf M}$. 
The action constructed from this Lagrangian is invariant under supersymmetry transformation rule given by
\beal{
\Delta A_{\bf M} =  \bar\epsilon \Gamma_{\bf M} \lambda,\quad
\Delta\lambda = \half  F_{{\bf M}{\bf N}} \Gamma^{{\bf M}{\bf N}} \epsilon
\label{10dsusytr}
}
where $\epsilon$ is a supersymmetry parameter of Majorana-Weyl fermion satisfying $\hat\Gamma_{10}\epsilon = \epsilon$ and $C_{10}\bar\epsilon^T=-\epsilon$.
The supersymmetry current is obtained as
\beal{
\ol{S^{\mbf P}}={1\over g_{10}^2}\Tr \biggl[\ol{\lambda } \half  F_{{\mbf M}{\mbf N}} \Gamma^{\mbf P}\Gamma^{{\mbf M}{\mbf N}} \biggr], \quad 
{ S^{\mbf P}}={1\over g_{10}^2}\Tr \biggl[- \half  F_{{\mbf M}{\mbf N}}\Gamma^{{\mbf M}{\mbf N}} \Gamma^{\mbf P}\lambda \biggr] 
\label{10dsusycurrent}
}
where the SUSY current satisfies
$\ol{ S^{\mbf P}} \epsilon= \ol{\epsilon} S^\mbf P$.

To compute the supersymmetry algebra of this theory, we compute variation of the SUSY current under supersymmetry transformation. 
\beal{
2g_{10}^2\Delta \ol{S^{\mbf P}} =& 
\Tr[\Delta\bar\lambda  F_{\mbf M\mbf N}\Gamma^\mbf P\Gamma^{\mbf M\mbf N}+2\bar\lambda D_\mbf M\Delta  A_\mbf N \Gamma^\mbf P\Gamma^{\mbf M\mbf N}]. 
}
The 1st term can be calculated as 
\beal{
\Tr[\Delta\bar\lambda  F_{\mbf M\mbf N}\Gamma^\mbf P\Gamma^{\mbf M\mbf N}]=&-\half \Tr[ F_{\mbf Q\mbf R} F_{\mbf M\mbf N}]\bar \epsilon\Gamma^{\mbf R\mbf Q\mbf P\mbf M\mbf N}-4\Tr[ F^{\mbf P\mbf M} F_{\mbf M\mbf N}]\bar\epsilon\Gamma^\mbf N-\Tr[ F^{\mbf M\mbf N} F_{\mbf M\mbf N}]\bar\epsilon\Gamma^\mbf P
}
The 2nd term is calculated as follows. 
\beal{
\Tr[2\bar\lambda D_\mbf M\Delta  A_\mbf N \Gamma^\mbf P\Gamma^{\mbf M\mbf N}] 
=&2\Tr[\bar\epsilon\Gamma_\mbf ND_\mbf M\lambda \bar\lambda\Gamma^\mbf P\Gamma^{\mbf M\mbf N}] \nn
=&{-1\over8}\bigg(\Tr[\bar\lambda\Gamma^{\mbf Q}D_\mbf M\lambda]\bar\epsilon\Gamma_\mbf N\Gamma_{\mbf Q}\Gamma^\mbf P\Gamma^{\mbf M\mbf N}+{1\over3!}\Tr[\bar\lambda\Gamma^{\mbf Q\mbf R\mbf S}D_\mbf M\lambda]\bar\epsilon\Gamma_\mbf N\Gamma_{\mbf S\mbf R\mbf Q}\Gamma^\mbf P\Gamma^{\mbf M\mbf N}\nn
&+{1\over2\cdot5!}\Tr[\bar\lambda\Gamma^{\mbf Q\mbf R\mbf S\mbf T\mbf U}D_\mbf M\lambda]\bar\epsilon\Gamma_\mbf N\Gamma_{\mbf U\mbf T\mbf S\mbf R\mbf Q}\Gamma^\mbf P\Gamma^{\mbf M\mbf N}\bigg),
\label{2nd}
}
in which we used ten dimensional Fierz identity 
\beal{
\chi\bar\psi={-1\over2^4}\bigg(\bar\psi\Gamma^\mbf M\chi\Gamma_\mbf M+{1\over3!}\bar\psi\Gamma^{\mbf M\mbf N\mbf P}\chi\Gamma_{\mbf P\mbf N\mbf M}+{1\over2\cdot5!}\bar\psi\Gamma^{\mbf M\mbf N\mbf P\mbf Q\mbf R}\chi\Gamma_{\mbf R\mbf Q\mbf P\mbf N\mbf M}\bigg){1-(-)^\psi\hat\Gamma_{10}\over2}
}
where $\psi,\chi$ are Weyl fermions of the same chirality and we denote the chirality of $\psi$ by $(-)^\psi$.
By using the equation of motion of gaugino $\Gamma^\mbf M D_\mbf M\lambda=0$ and a formula 
\be 
\Gamma^{\mbf M_1\cdots \mbf M_5}\psi\bar\chi\Gamma_{\mbf M_1\cdots \mbf M_5}=0
\ee
where $\psi$ and $\chi$ are Weyl fermion with the same chirality, 
the above can be simplified as
\beal{
\eqref{2nd}=& \Tr[\bar\lambda\Gamma_{\mbf M_1}D_{\mbf M_2}\lambda]\bar\epsilon\Gamma^{\mbf M_2\mbf M_1}{}^\mbf P+2\Tr[\bar\lambda\Gamma^{\mbf P}D_{\mbf M}\lambda]\bar\epsilon\Gamma^{\mbf M}
+\Tr[\bar\lambda\Gamma^\mbf P{}_{\mbf M_1\mbf M_2}D_{\mbf M_3}\lambda]\bar\epsilon\Gamma^{\mbf M_3\mbf M_2\mbf M_1}.
}
Summing up these terms we find%
\footnote{ 
Our result in the fermionic part is different from that in \cite{Osborn:1979tq}. One of the reasons is that the stress tensor given in \cite{Osborn:1979tq} is not a symmetric one in the fermionic part. However the argument there does not need modification since the fermionic part was neglected in other parts of that paper. 
}
\beal{
2g_{10}^2\Delta \ol{ S_\mbf P } =&
 -4g_{10}^2 T_{\mbf P}{}_\mbf M\bar\epsilon\Gamma^\mbf M
-\half \Tr[ F_{\mbf Q\mbf R} F_{\mbf M\mbf N}]\bar \epsilon\Gamma_\mbf P{}^{\mbf R\mbf Q\mbf M\mbf N}
-{1\over4}\partial^{\mbf M_3} \Tr[\bar\lambda\Gamma_{\mbf M_1\mbf M_2\mbf M_3}\lambda]\bar\epsilon\Gamma_\mbf P{}^{\mbf M_2\mbf M_1} \nn
&+\half \partial_{\mbf M_3}\Tr[\bar\lambda\Gamma_{\mbf P\mbf M_1\mbf M_2}\lambda]\bar\epsilon\Gamma^{\mbf M_3\mbf M_2\mbf M_1}
-\half \partial^\mbf N\Tr[\bar\lambda \Gamma_{\mbf P\mbf M\mbf N}\lambda]\bar\epsilon\Gamma^{\mbf M}
\label{deltasusycurrent}
}
where $T_{\mbf P\mbf M}$ is the stress tensor given by
\beal{ 
T_{\mbf M\mbf P}=&{1\over 4 g_{10}^2}\bigg(4\Tr[ F_\mbf P{}^\mbf N F_{\mbf N\mbf M}]+\eta_{\mbf M\mbf P}\Tr[ F_{\mbf M\mbf N} F^{\mbf M\mbf N}]-2\Tr[\bar\lambda\Gamma_{(\mbf M}D_{\mbf P)}\lambda] \bigg)
}
and we used 
$
\bar\lambda\Gamma_{\mbf M_1\mbf M_2\mbf M_3}D_\mbf M\lambda
=\half D_\mbf M(\bar\lambda\Gamma_{\mbf M_1\mbf M_2\mbf M_3}\lambda), 
$
and $X_{(A}Y_{B)}:= \half (X_{A}Y_{B}+X_{B}Y_{A})$.

The supercharge is defined by 
\beal{
Q=\int d^9x S^0. \quad 
}
Under the standard convention of canonical formalism, 
it can be shown that 
\be 
\Delta \cO = [-i \bar\epsilon Q, \cO]
\label{deltaO}
\ee
for a gauge invariant operator $\cO$ and a canonical bracket. 
Although it is not difficult to show this computationally, 
it needs a little careful argument to justify this, as we shall do below. 
The canonical momentum of the gaugino is computed as 
\beal{
\Pi_\lambda = {\partial \cL \over \partial (\partial_0 \lambda)} ={1\over2 g_{10}^2} [-\bar\lambda\Gamma^0 ]. 
\label{10dcanonicalmomentagaugino} 
}
Under the canonical commutation relation 
$
[\Pi_\lambda, \lambda] = i \delta, 
$
where $\delta$ is the unit {\it matrix} in terms of implicit space, gauge and spinor indices, 
one can easily show that 
\be 
\Delta \lambda = [-i \bar\epsilon Q, \lambda]. 
\ee 
On the other hand, 
the canonical momentum of the gauge field is computed as 
\beal{
\Pi_{ A_\mbf M} = {\partial \cL \over \partial (\partial_0  A_\mbf M)} ={1\over g_{10}^2}  F^{0\mbf M}, \quad 
\label{10dcanonicalmomentagauge} 
}
which has a vanishing component for time direction as ordinary Yang-Mills theory. 
This suggests that there is no kinetic term of the time component of the gauge field in the (off-shell) Lagrangian and the system is constrained by saddle point equation thereof, which is given by $D_M\Pi^M=0$, where $M$ runs the space directions. 
This requires us to choose a set of dynamical (or canonical) variables to quantize the system. We naturally choose it as the gauge fields of the space directions. 
Then the canonical commutation relation is 
$
[\Pi_{ A_M},  A_N] = i\delta^M_N \delta. 
$
By using this it is not difficult to show that 
\be 
\Delta  A_M = [-i \bar\epsilon Q,  A_M]. 
\label{deltagauge}
\ee
We stress that the SUSY variation \eqref{10dsusytr} is reproduced for the dynamical gauge fields ($ A_M$) and not for the auxiliary one ($ A_0$).%
\footnote{ 
This standpoint may be different from one argued in \cite{Popescu:2001rf}, where SUSY algebra of a general four dimensional $\cN=2$ SYM of an $\cN=2$ vector multiplet was studied. It seems there that the SUSY variation of all the components of the gauge fields was reproduced in Appendix D, which may be incorrect for that of the auxiliary gauge field.     
}
This argument is consistent with the fact that the SUSY variation of supercurrent derived in \eqref{deltasusycurrent} is an on-shell relation. 
One may ask that there will be another constraint by fixing gauge symmetry which every Yang-Mills theory possesses, in which case one has to use not the canonical bracket but a Dirac one for \eqref{deltagauge} in order to be consistent with the gauge fixing. 
This should be the case though we still claim that \eqref{deltaO} holds for a canonical bracket. 
The argument is as follows. 
When one fixes gauge symmetry, the initial supersymmetry transformation is not consistent with the fixed gauge in general. 
One can modify the SUSY transformation so as to be consistent with the gauge fixing by combining gauge transformation.  
Then the right-hand side of \eqref{deltagauge} replaced by the Dirac bracket will reproduce the modified SUSY transformation for the gauge fields. This suggests that the modified SUSY transformation for a gauge invariant operator should agree with the initial one because the modification is given by a gauge transformation. 
Thus one has only to use a canonical bracket and do not need to use a Dirac one in \eqref{deltaO}.

As a result, by using \eqref{deltasusycurrent} and \eqref{deltaO}, algebra between the supercurrent and supercharge in SYM in ten dimensions (local form of SUSY algebra) is given by 
\beal{ 
\{Q,\ol{ S_\mbf P }\} =&
-2i T_{\mbf P}{}_\mbf M\Gamma^\mbf M
+ J_{\mbf P\mbf M} \Gamma^\mbf M 
+ J_{\mbf P\mbf M_1\mbf M_2 \mbf M_3} \Gamma^{\mbf M_3\mbf M_2\mbf M_1}+i C_{\mbf M_1\mbf M_2 } \Gamma_\mbf P{}^{\mbf M_2\mbf M_1} \nn
&  +J_\mbf P{}^{\mbf M_5\mbf M_4\mbf M_3\mbf M_2\mbf M_1}\Gamma_{\mbf M_1\mbf M_2\mbf M_3\mbf M_4\mbf M_5}
\label{10dlocalsusyalgebra}
}
where we define
\beal{ 
J_{\mbf P\mbf M}=& {-i \over 4g_{10}^2}\partial^\mbf N\Tr[\bar\lambda \Gamma_{\mbf P\mbf M\mbf N}\lambda], \\
C_{\mbf M_1\mbf M_2 } =&-{1\over8g_{10}^2} \partial^{\mbf M_3} \Tr[\bar\lambda\Gamma_{\mbf M_1\mbf M_2\mbf M_3}\lambda], \\
J_{\mbf P\mbf M_1\mbf M_2 \mbf M_3}=&{i\over 4g_{10}^2} \partial_{\mbf M_3}\Tr[\bar\lambda\Gamma_{\mbf P\mbf M_1\mbf M_2}\lambda], \\
J_\mbf P{}^{\mbf M_5\mbf M_4\mbf M_3\mbf M_2\mbf M_1}=&-{i\over 4g_{10}^2}\Tr[ F_{\mbf Q\mbf R} F_{\mbf M\mbf N}]\varepsilon_\mbf P{}^{\mbf Q\mbf R\mbf M\mbf N\mbf M_5\mbf M_4\mbf M_3\mbf M_2\mbf M_1},
}
with $\varepsilon_{01\cdots9}=1$. 
Note that the contributions of fermions are total derivative terms.
Especially we obtain supersymmetry algebra in ten dimensional SYM as
\beal{
\{Q,\ol{Q}\} =&
-2i P_{\mbf M}\Gamma^\mbf M +Z_{\mbf M} \Gamma^\mbf M + Z_{\mbf M_1\mbf M_2 \mbf M_3}\Gamma^{\mbf M_3\mbf M_2 \mbf M_1}  + Z_{\mbf M_5\mbf M_4\mbf M_3\mbf M_2\mbf M_1}\Gamma^{\mbf M_1\mbf M_2\mbf M_3\mbf M_4\mbf M_5}
\label{10dsusyalgebra}
}
where we used $\partial^{\mbf M_3} \Tr[\bar\lambda\Gamma_{\mbf M_1\mbf M_2\mbf M_3}\lambda]\bar\epsilon\Gamma_0{}^{\mbf M_2\mbf M_1}=0$ on shell, and we set  
\bes{ 
P^\mbf M=&\int d^9x T^{0\mbf M}, \quad Z^\mbf M= \int d^9x J^{0\mbf M}, \\ 
Z^{\mbf M_1\mbf M_2 \mbf M_3}=&\int d^9x J^{0\mbf M_1\mbf M_2 \mbf M_3}, \quad 
Z^{\mbf M_5\mbf M_4\mbf M_3\mbf M_2\mbf M_1}=\int d^9x J^{0\mbf M_5\mbf M_4\mbf M_3\mbf M_2\mbf M_1}.
}

\section{Superalgebra in 6d SYM}
\label{6d}
In this section we compute supersymmetry algebra of six dimensional SYM  with eight and sixteen supercharges. 
We derive results of maximally SYM in six dimensions by dimensional reduction of ten dimensional one obtained in the previous section. 

\subsection{Vector multiplet} 
\label{6dVM}
First we consider a vector multiplet. 
This theory has $SU(2)\simeq Sp(1)$ global symmetry.  
The bosonic field contents are a gauge field $ A_M$, $M=0,1, \cdots, 5$ and the SU(2) triplet auxiliary fields 
$D^A\!_B$ which satisfy $(D^A\!_B)^\dagger= D^B\!_A$, $D^A\!_A=0$. 
The super partner $\lambda^A$ is a Sp(1)-Majorana Weyl fermion satisfying 
\be
\hat\Gamma \lambda^A = + \lambda^A, \quad 
\varepsilon^{AB} C_6 (\ol{\lambda^B})^T =\lambda^A
\label{6dMW}
\ee 
where $\varepsilon_{12}=\varepsilon^{12}=1$, and $\hat\Gamma$ and $C_6$ are a chirality matrix and a charge conjugation in six dimensions, respectively, defined by
\beal{
\hat \Gamma=  \Gamma_{012345}, 
\quad C_6=\Gamma_{035}.
}
See Appendix\ref{convention} for more details on our convention in six dimensions. 
In this convention, the supersymmetric Lagrangian reads
\beal{ 
{\cal L}_V
=&{1\over g_6^2}
\Tr \biggl[{1\over 4}  F_{MN}  F^{MN} + \half \ol{\lambda^A} \Gamma^M[{D_M},\lambda^A] 
+\half D^A{}_B D^B\!_A\biggr] 
\label{6dn1Lagrangian}
}
where $ F_{MN}= [D_M,D_N]$, $D_M= \partial_M+ A_M$.
The supersymmeric transformation rule is 
\bes{
\Delta  A_M =& \ol{{\epsilon}^A} \Gamma_M \lambda^A,\\
\Delta \lambda^A =& \half   F_{MN} \Gamma^{MN} {\epsilon}^A +\alpha D^A\!_B {\epsilon}^B, \\
\Delta D^A\!_B =&\alpha ( D_M \ol{\lambda^B} \Gamma^M {\epsilon}^A - \half \delta^A_B D_M \ol{\lambda^C} \Gamma^M {\epsilon}^C), 
\label{6dsusytr}
}
where $\epsilon^A$ is also a symplectic-Majorana Weyl fermion such that $\hat\Gamma\epsilon^A=\epsilon^A$ and $\varepsilon^{AB} C_6 (\ol{\epsilon^B})^T = -\epsilon^A$. 
Thus the type of SUSY is (1,0). 
$\alpha$ is arbitrary parameter and thus one can set to zero as long as one considers only vector multiplet due to the fact that the auxiliary field can be integrated out to be zero. Once one introduces coupling to a hyper multiplet, which is done in the next subsection, $\alpha$ is uniquely determined as $\alpha = \sqrt 2$. 
The supersymmetry current of this theory is computed in the same way as in ten dimensions.
\beal{
\overline{S^A_P} = {1\over g_6^2}\Tr \biggl[{1\over2} \ol{\lambda^A}  F_{MN} \Gamma_P\Gamma^{MN} \bigg], \quad S^A_P = {1\over g_6^2}\Tr \biggl[-{1\over2}   F_{MN}\Gamma^{MN} \Gamma_P {\lambda^A} \bigg],
\label{6dsusycurrent}
}
where they are determined so as to satisfy 
$\ol{\epsilon^A}S^A_P = \ol{S^A_P}\epsilon^A$. 

Let us compute the SUSY algebra of $\cN=1$ SYM consisting of a vector multiplet.  
\beal{
2g_{6}^2\Delta_\epsilon \overline{S^A_P} =& 
\Tr[\Delta\overline{\lambda^A}  F_{MN}\Gamma_P\Gamma^{MN}+2\overline{\lambda^A} D_M\Delta  A_N\Gamma_P\Gamma^{MN}]. 
}
The 1st term can be calculated as 
\beal{
\Tr[\Delta\overline{\lambda^A}  F_{MN}\Gamma_P\Gamma^{MN}]=&-\half \Tr[ F_{QR} F_{MN}]\ol{\epsilon^A}\Gamma^{RQ}{}_P{}^{MN}-4\Tr[ F_P{}^M F_{MN}]\ol{\epsilon^A}\Gamma^N-\Tr[ F^{MN} F_{MN}]\ol{\epsilon^A}\Gamma_P\nn
&-\alpha\Tr[(D^A\!_B)^\dagger  F_{MN}]\ol{\epsilon^B}(\Gamma_P{}^{MN}+\delta^M_P\Gamma^N-\delta^N_P\Gamma^M).
}
The 2nd term is calculated as follows. 
\beal{
\Tr[2\overline{\lambda^A} D_M\Delta  A_N\Gamma_P\Gamma^{MN}] 
=& 2\Tr[\ol{\epsilon^B}\Gamma_ND_M\lambda^B \ol{\lambda^A}\Gamma_P\Gamma^{MN}]\nn
=&{-1\over2}\Tr[\ol{\lambda^A}\Gamma^{M_1}D_M\lambda^B]\ol{\epsilon^B}\Gamma_N\Gamma_{M_1}\Gamma_P\Gamma^{MN} \nn
&-{1\over24}\Tr[\ol{\lambda^A}\Gamma^{M_1M_2M_3}D_M\lambda^B]\ol{\epsilon^B}\Gamma_N\Gamma_{M_3M_2M_1}\Gamma_P\Gamma^{MN}
}
where we used a Fierz identity 
\beal{
\chi\bar\psi={-1\over2^2}\bigg(\bar\psi\Gamma^M\chi\Gamma_M+{1\over3!\cdot 2}\bar\psi\Gamma^{MNP}\chi\Gamma_{PNM}\bigg){1-(-)^\psi\hat\Gamma\over2}
}
for Weyl fermions $\psi,\chi$ with the same chirality.
By using 
\beal{
\Gamma^{M_1M_2M_3}\psi\bar\chi\Gamma_{M_1M_2M_3}=0
}
where $\psi$ and $\chi$ are Weyl fermions with  the same chirality, and 
\beal{
\ol{\lambda^A}\Gamma_PD_M\lambda^B=\half D_M(\ol{\lambda^A}\Gamma_P\lambda_B)+\half\delta^B_A\ol{\lambda^C}\Gamma_PD_M\lambda^C
}
we find  
\beal{
\Tr[2\overline{\lambda^A} D_M\Delta  A_N\Gamma_P\Gamma^{MN}] =&2\partial_M\Tr[\ol{\lambda^A}\Gamma_{P}\lambda^B]\ol{\epsilon^B}\Gamma^{M}+2\Tr[\ol{\lambda^C}\Gamma_{P}D_{M}\lambda^C]\ol{\epsilon^A}\Gamma^{M}
}
where we also used the equation of motion of gaugino. 
Collecting these we obtain 
\beal{
2g_{6}^2\Delta \ol{S^A_P} =&
-\half \Tr[ F_{QR} F_{MN}]\ol{\epsilon^A}\Gamma^{RQ}{}_P{}^{MN}-4\Tr[ F_P{}^M F_{MN}]\ol{\epsilon^A}\Gamma^N-\Tr[ F^{MN} F_{MN}]\ol{\epsilon^A}\Gamma_P \nn
&-\alpha\Tr[(D^A\!_B)^\dagger  F_{MN}]\ol{\epsilon^B}(\Gamma_P{}^{MN}+2\delta^M_P\Gamma^N)\nn
&+2\partial_M\Tr[\ol{\lambda^A}\Gamma_{P}\lambda^B]\ol{\epsilon^B}\Gamma^{M}+2\Tr[\ol{\lambda^C}\Gamma_{P}D_{M}\lambda^C]\ol{\epsilon^A}\Gamma^{M}.
}
Integrating out the auxiliary field gives $D^A\!_B=0$. 
Then
\beal{
2g_{6}^2\Delta \ol{S^A_P}
=& -4g_{6}^2 T_{PM}\ol{\epsilon^A}\Gamma^M
-\half \Tr[ F_{QR} F_{MN}]\ol{\epsilon^A}\varepsilon_P{}^{QRMNL}\Gamma_L \nn
&+2\partial_M\Tr[\ol{\lambda^A}\Gamma_{P}\lambda^B]\ol{\epsilon^B}\Gamma^{M}+2\Tr[\ol{\lambda^C}\Gamma_{[P}D_{M]}\lambda^C]\ol{\epsilon^A}\Gamma^{M}
}
where $\varepsilon_{012345}=1$, $T_{PM}$ is the stress tensor on shell given by
\beal{ 
T_{MP}
=& {1\over 4g_6^2} 
\Tr \big[ g_{MP}  F_{QN}  F^{QN} +4  F_{P}{}^N  F_{NM} -2\ol{\lambda^A} \Gamma_{(M} D_{P)}\lambda^A   \big]. 
}
A supercharge with Sp(1) index is defined by 
\beal{
Q^A=\int d^5x S^{0A}. 
\label{6dsusycharge}
}
In the same argument given in \S\ref{10d}, 
one can show that $\Delta \cO = [-i \ol{\epsilon^A} Q^A, \cO]$ for a gauge invariant operator $\cO$.  
Then local form of supersymmetry algebra of SYM in six dimensions is determined as 
\beal{
\{Q^B,\ol{S^A_P}\} =& 
(-2i \delta^B_A T_{PM} +\delta^B_A J_{PM}  +\delta^B_A J'_{PM}+J^B_A{}_{PM})\Gamma^{M} 
\label{6dlocalsusy}
}
where 
\beal{
J_{PM}=& -{i\over 4g_6^2} \Tr[ F_{QR} F_{LN}]\varepsilon_{PM}{}^{QRLN}, \\
J'_{PM}=&-{i\over2g_6^2}  \partial^N\Tr[\ol{\lambda^C}\Gamma_{PMN}\lambda^C], \\
J^B_A{}_{PM}=& {i\over g_6^2} \partial_M\Tr[\ol{\lambda^A}\Gamma_{P}\lambda^B].
}
There are several comments. 
Firstly as in ten dimensional case the contributions of fermions are given by total derivative terms. 
Secondly the terms in the right-hand side are all conserved, which is consistent with the fact that the SUSY current in the left-hand side is conserved. Especially for $J_{PM}, J'_{PM}$, these are off-shell divergenceless.  
These anti-symmetric tensors are not distinguishable in the algebra \eqref{6dlocalsusy}. 
There also exists non-R symmetric tensor $J^B_A{}_{PM}$.   
Those tensors are so-called brane currents \cite{Dumitrescu:2011iu}, which describes extended BPS objects in the theory.  

One might ask whether total derivative terms of fermions appearing in the superalgebra are truly physical or not,
since they may be absorbed by an improvement transformation preserving SUSY.%
\footnote{The author would like to thank the referee for raising this question.} 
A general study of this was done in four dimensions by using superfield formalism \cite{Dumitrescu:2011iu}. 
As a result an improvement transformation keeping SUSY including operators with spin not more than one was determined.%
\footnote{ Existence of an improvement transformation to kill a total derivative term is not sufficient to decide the term as unphysical. 
To decide so, it also requires fields constructing the term to fall off fast enough at spatial infinity.  } 
And a general supercurrent multiplet called ${\cal S}$-multiplet was classified into several irreducible supercurrent multiplets by whether there exists an improvement transformation to kill a submultiplet inside the ${\cal S}$-multiplet. 
To perform this kind of general analysis of supercurrent in the current case, it is important to develop superfield formalism in six dimensions which can determine an improvement transformation including higher spin operators.
We leave these problems to future work. 

Volume integration of both sides of \eqref{6dlocalsusy} leads to supersymmetry algebra of six dimensional SYM theory as 
\beal{
\{Q^B,\ol{Q^A}\} =&
(\delta^B_A (-2 i P_{M} +Z_M+ Z_M') + Z_A^B{}_M )\Gamma^M 
\label{6dsusyalgebra}
}
where we set 
\beal{
&P^M=\int d^5x T^{0M}, \quad Z_M=\int d^5x J{}^0\!_{M}, \quad 
Z'_M= \int d^5x J'{}^0\!_{M}, \quad 
Z^B_A{}_M= \int d^5x  J^B_A{}^0\!_{M}. 
}
$Z_M, Z_M',Z^B_A{}_M$ are {\it brane charges} corresponding to the brane currents mentioned above.

\subsection{Inclusion of a hyper multiplet} 
\label{6dHM}

In this subsection we determine supersymmetry algebra of six dimensional SYM including a hyper multiplet. 
Extension to a multiple case is straightforward. 
A hyper multiplet consists of two complex scalar fields $q^A$, $A=1,2$, and a chiral fermion $\psi$ which has the opposite chirality to that of gaugino to interact therewith: 
$
\hat\Gamma\psi = -\psi. 
$
We consider a case where the hyper multiplet is in the fundamental representation of the gauge group for notational simplicity. Generalization to other representation can be easily done. 
The supersymmetric Lagrangian of the hyper multiplet is given by
\be
\cL_{H}{=} -D_M (q^A)^\dagger D^M q^A + \half \ol{\psi}\slash\!\!\!\! D \psi  
+ \varepsilon^{AB} (q^A)^\dagger \ol{\lambda^B} \psi 
- \varepsilon_{AB} \bar\psi \lambda^A q^B 
+ \sqrt2 (q^A)^\dagger D^A\!_B q^B 
\label{6dhyperlagrangian}
\ee
and the supersymmetry transformation is determined as 
\beal{
\Delta q^A =& \varepsilon^{AB} \ol{\epsilon^B}\psi, \quad  
\Delta (q^A)^\dagger =\varepsilon_{AB} \bar\psi \epsilon^B, \\
\Delta \psi =& 2 \varepsilon_{BA} \Gamma^M  \epsilon^B D_M q^A, \quad 
\Delta \bar\psi=  -2\varepsilon^{BA} \ol{\epsilon^B} \Gamma^M D_M (q^A)^\dagger.
\label{6dsusytrhyper}
}
The variation of the action of a hyper multiplet under the SUSY transformation is computed as 
\beal{ 
\Delta\(\int d^6x \cL_H\)=\int d^6x \ol{S^A_M}_{\text{hyp}} \partial_M\epsilon^A
}
where 
\be
\ol{S^A_P}_{\text{hyp}} = \varepsilon_{AB} \ol{\psi}\Gamma_P \Gamma^N   D_N q^B-(D_N q^A)^\dagger \psi^T C_6\Gamma_P \Gamma^N -2 (q^A)^\dagger   \ol{\lambda^B} \Gamma_P  q^B + (q^B)^\dagger \ol{\lambda^A} \Gamma_P q^B.
\ee
Thus the supercurrent is given by
\beal{
\ol{S^A_P}=&{1\over 2g_6^2}\Tr[\ol{\lambda^A} F_{MN}\Gamma_P\Gamma^{MN}] +\varepsilon_{AB} \ol{\psi}\Gamma_P  \Gamma^N   D_N q^B-(D_N q^A)^\dagger \psi^T C_6\Gamma_P \Gamma^N \nn 
&-2 (q^A)^\dagger   \ol{\lambda^B} \Gamma_P q^B + (q^B)^\dagger \ol{\lambda^A} \Gamma_P q^B. 
\label{6dsusycurrenthyper}
}
Note that $S^A_P$ can be determined by using $\ol{\epsilon^A}S^A_P = \ol{S^A_P}\epsilon^A$.

We can show that the supersymmetry current \eqref{6dsusycurrenthyper} is conserved: $\partial^M \ol{S^A_M}=0$ on shell.
To show this, we need equations of motion of the gauge multiplet 
\beal{ 
&{1\over g_6^2} D_N F^{NM}=-{1\over g_6^2}\ol{\lambda^A}\Gamma^M\lambda^A+D^M q^A (q^A)^\dagger  -q^A D^M(q^A)^\dagger -\half\psi^T(\Gamma^M)^T \bar\psi^T, \\
&{1\over g_6^2} D_M \ol{\lambda^A}\Gamma^M= -(\varepsilon_{AB}q^B\bar\psi+\psi^T C_6(q^A)^\dagger), \\
&D^A\!_B = -\sqrt 2 g_6^2 (q^A(q^B)^\dagger - \half\delta^A_B q^C(q^C)^\dagger),
}
and those of the hyper multiplet   
\beal{ 
&D^2 q^A +\varepsilon^{AB}\ol{\lambda^B}\psi+\sqrt2 D^A\!_B q^B=0, \\
&\half \sla D\psi + \varepsilon_{AB}\lambda^B q^A=0,\quad -\half D_M\bar\psi \Gamma^M+ \varepsilon^{AB}(q^A)^\dagger\ol{\lambda^B}=0. 
}
We also need to employ another Fierz rearrangement
\beal{
\chi\bar\psi={-1\over 4}(\bar\psi\chi+{1\over2}\bar\psi\Gamma^{MN}\chi \Gamma_{NM}){1-(-)^\psi\hat\Gamma\over2}
}
where $\psi, \chi$ are Weyl fermions with different chirality, 
and a formula 
\be
\Tr[\ol{\lambda^A}\Gamma^M (\ol{\lambda^B}\Gamma_M\lambda^B)]=0.
\ee

Let us determine supersymmetry algebra in six dimensional SYM theory including a hyper multiplet. 
As seen from the equations of motion above, 
it is complicated to determine SUSY algebra including fermionic sector, thus we neglect the fermionic part in this paper, which we leave to future work. 
The variation of supercurrent under the supersymmetry transformation is computed as follows.   
\beal{
\Delta \ol{ S^A_P}
=&  -2 T_{PM}\ol{\epsilon^A}\Gamma^M
+{1\over4g_6^2} \Tr[ F_{QR} F_{MN}]\ol{\epsilon^A}\Gamma_P{}^{QRMN}  \nn
&-4\partial_M [(q^A)^\dagger D_N q^B-\half \delta^B_A(q^C)^\dagger D_N q^C]\ol{\epsilon^B}\Gamma_P{}^{MN}
\label{6ddeltasusycurrent}
}
where the stress tensor of the bosonic fields is given by
\beal{ 
T_{MP}
=& {1\over 4g_6^2} 
\Tr \big[ g_{MP} ( F_{QN}  F^{QN}+\half D^A\!_B D^B\!_A) +4  F_{P}{}^N  F_{NM} \big] \nn
&+2D_{(M}(q^A)^\dagger D_{P)} q^A -g_{MP}\partial_N((q^A)^\dagger D^N q^A).
}
Note that the quartic terms of the complex scalar fields vanish, which is required from consistency with conservation of the supercurrent in the left hand side of the superalgebra. 
As in the previous sections we can show that $\Delta \cO = [-i\ol{\epsilon^A} Q^A, \cO]$. 
Thus we obtain
local form of supersymmetry algebra of six dimensional SYM including a hyper multiplet.  
\beal{ 
\{Q^B,\ol{ S^A_P}\}
=& \delta^B_A (-2i T_{PM}\Gamma^M+ J_{PM}\Gamma^M)
+C^B_A{}_{PQRS}\Gamma^{SRQ}
\label{6dlocalsusyhyper}
}
where 
\beal{
C^B_A{}^{PQRS}=&{2i \over 3}\partial_M [(q^A)^\dagger D_N q^B-\half \delta^B_A(q^C)^\dagger D_N q^C]\varepsilon^{PMNQRS}.
}
Thus supersymmetry algebra in six dimensional SYM including a hyper multiplet is obtained as 
\beal{
\{Q^B,\ol{Q^A}\} =&
\delta^B_A(-2iP_{M} + Z_M)\Gamma^M + Y_A^B{}_{MNP} \Gamma^{PNM}
\label{6dsusyalgebrahyper}
}
where we set 
\beal{
Y_A^B{}^{QRS} = \int d^5x \; C^B_A{}^{0QRS}.
}

\subsection{6d $\cN=2$ superalgebra } 
\label{6dn2}

In this section we determine supersymmetry algebra of $\cN=2$ SYM by performing dimensional reduction from that of ten dimensional SYM, which was computed in \S\ref{10d}. 
$\cN=2$ SYM in six dimensions is constructed by a pair of vector multiplet and hyper multiplet in the adjoint representation. 
Thus six dimensional $\cN=2$ SYM Lagrangian is given by 
addition of the Lagrangians of those multiplets, which were derived in the previous subsections. 
\beal{
{\cal L}^{\cN = 2}
=&{1\over g_6^2}
\Tr \biggl[{1\over 4}  F_{MN}  F^{MN} + \half \ol{\lambda^A} \sla{D}\lambda^A 
+\half D^A\!_B D^B\!_A + \sqrt2 (q^A)^\dagger [D^A\!_B, q^B] \nn
& -D_M (q^A)^\dagger D^M q^A + \half \ol{\psi}\sla D\psi 
+ \varepsilon^{AB} (q^A)^\dagger [\ol{\lambda^B}, \psi ]
+\varepsilon_{AB} \bar\psi[\lambda^B, q^A]
 \biggr]. 
}
Note that the coupling constants of the vector multiplet and the hyper multiplet are the same. 
Integrating out the auxiliary field results in 
\beal{
{\cal L}^{\cN = 2}
=&{1\over g_6^2}
\Tr \biggl[{1\over 4}  F_{MN}  F^{MN} + \half \ol{\lambda^A} \sla {D} \lambda^A 
- \half D'^A\!_B D'^B\!_A \nn
& -D_M (q^A)^\dagger D^M q^A + \half \ol{\psi}\sla D \psi 
+ \varepsilon^{AB} (q^A)^\dagger [\ol{\lambda^B}, \psi ]
+\varepsilon_{AB} \bar\psi[\lambda^B, q^A]
 \biggr]
\label{6dn2Lagrangian}
}
where 
\be
D'^A\!_B = -\sqrt 2 \([q^A, (q^B)^\dagger] -\half \delta^A_B[q^C, (q^C)^\dagger]\).
\ee

In order to determine $\cN=2$ supersymmetry transformation and show that the Lagrangian \eqref{6dn2Lagrangian} has sixteen maximal supersymmetry, we perform dimensional reduction for the SYM Lagrangian in ten dimensions. 
We compactify four directions $x^{m+5}$, where  $m=1,2,3,4$. Then $X_m =-i  A_{m+5}$ become four real scalar fields in six dimensions. 
We decompose the SO(1,9) gamma matrices denoted by $\Gamma^{(10)}_{M}$ as 
\bea
\Gamma^{(10)}_{M}=\Gamma_{ M} \otimes \mbf 1, \quad 
\Gamma^{(10)}_{m+5}=\hat\Gamma \otimes \gamma_m \,
\label{10dgamma}
\eea
where $\Gamma_{ M}\, ({M}=0,\cdots, 5)$ are SO(1,5) gamma matrices and $\gamma_m$ are SO(4) gamma matrices, which we realize by a chiral expression  
\beal{
\gamma_m= 
\begin{pmatrix}
 0 & \bar\sigma_m \\
 \sigma_m & 0
\end{pmatrix}
} 
where $\bar\sigma_i=\sigma_i$ $(i=1,2,3)$ are Pauli matrices and $\bar\sigma_4=-\sigma_4=i$. 
Then the chirality matrix and charge conjugation matrix in ten dimensions are computed as 
\beal{
\hat\Gamma_{10} =&\hat\Gamma\otimes
\left(\begin{array}{cc}
1 & 0\\
0 & -1\\
\end{array}
\right) , \quad 
C_{10} =C_6 \otimes \left(
\begin{array}{cc}
i\sigma_2 & 0\\
0 & -i\sigma_2\\
\end{array}
\right).
} 
Therefore the Majorana-Weyl condition \eqref{10dMW} in ten dimensions reduces to  
\bea
\lambda = \left(
\begin{array}{c}
\lambda^A_{+}\\
\lambda^A_{-}
\end{array}
\right),
\quad \hat\Gamma \lambda^A_{\pm} = \pm\lambda^A_{\pm}, \quad
\lambda^A_\pm =\pm \varepsilon^{AB} C_6 \overline{\lambda^B_{\pm}}^T,
\eea
where $A=1,2$.  
This means that a ten-dimensional Majorana-Weyl fermion reduces to two sympletic-Majorana Weyl fermions $\lambda^A_\pm$ in six dimensions.
Then ten dimensional $\cN=1$ SYM Lagrangian reduces to 
\beal{
\cL^{\cN=2}=& \frac{1}{g_{6}^2} \Tr
\biggl[{1\over 4}  F_{{M}{N}}  F^{{M}{N}}
-\half D_{M} X_m D^{M} X^m
+{1\over4}[X_m,X_n][X^m,X^n]\nn
&+{1\over2}\ol{\lambda^A_+}\Gamma^{M} D_{M} \lambda^A_+
+{1\over2}\ol{\lambda^A_-} \Gamma^{M} D_{M} \lambda^A_-
+{1\over2} (\overline{\lambda^A_{+}} (\bar\sigma_{m})^A\!_B [i X^m, -\lambda_{-}^B]
+\overline{\lambda^A_{-}} (\sigma_{m})^A\!_B [i X^m, \lambda_{+}^B])
\biggl].
\label{6dn2Lagrangian2}
}
Note that this Lagrangian has manifest $SO(4) \simeq SU(2)\times SU(2)$ symmetry. 
This SO(4) symmetric Lagrangian \eqref{6dn2Lagrangian2} agrees with \eqref{6dn2Lagrangian} under the following identification.
\bes{ 
&\lambda^A_+ = \lambda^A, \quad \lambda^1_- = {1\over \sqrt2} \psi, \\
&X_1={(q^1)^\dagger -q^1 \over \sqrt2 i }, \; X_2={(q^1)^\dagger + q^1 \over - \sqrt2 }, \;
X_3={q^2-(q^2)^\dagger  \over \sqrt2 i }, \; X_4={q^2+(q^2)^\dagger \over \sqrt2  }.
}
The $\cN=2$ supersymmetry transformation rule 
boils down to 
\bes{
\Delta  A_{ {M}} =& \ol{\epsilon^A_+} \Gamma_{{M}}{\lambda^A_+}
+ \ol{\epsilon^A_-} \Gamma_{{M}}{\lambda^A_-}, \\
\Delta X_{m} 
=&i(\overline{\epsilon^A_{+}} (\bar\sigma_{m})^A\!_B \lambda_{-}^B
-\overline{\epsilon^A_{-}} (\sigma_{m})^A\!_B  \lambda_{+}^B),
 \\
\Delta\lambda^A_+=& \half  F_{ {M} {N}}\Gamma^{ {M} {N}}\epsilon^A_{+}-iD_{ {M}} X_m \Gamma^{ {M}}\bar\sigma^m{}^{A}\!_B\epsilon^B_{-}-{1\over 2} [X_m,X_n] \sigma^{mn}{}^{A}\!_B\epsilon^B_{+}, \\
\Delta\lambda^A_-=&  \half  F_{ {M} {N}}\Gamma^{ {M} {N}}\epsilon_{-}^A+i D_{ {M}} X_m \Gamma^{ {M}}\sigma^m{}^{A}\!_B\epsilon^B_{+}-
{1\over 2} [X_m,X_n] \bar\sigma^{mn}{}^{A}\!_B\epsilon^B_{-}.
\label{6dn2sym}
}
where the supersymmetry parameters $\epsilon^A_\pm$ satisfy
\bea
\hat\Gamma \epsilon^A_{\pm} = \pm\epsilon^A_{\pm}, \quad
\epsilon^A_\pm =\mp \varepsilon^{AB} C_6 \overline{\epsilon^B_{\pm}}^T.
\eea
Therefore the type of SUSY is (1,1). 

Performing dimensional reduction for ten dimensional SUSY current \eqref{10dsusycurrent}, we obtain two supersymmetry currents in six dimensions.  
\bes{
\ol{S_-{}^A_P}=&{1\over g_{6}^2} \Tr \bigg[\half \ol{\lambda_+^A} F_{{M}{N}} \Gamma^{P}\Gamma^{MN}-i \ol{\lambda_-^B}(\bar\sigma^{n})^B{}_{A}\Gamma^{P}\Gamma^{M} D_M X_n -\half \ol{\lambda_+^B}(\sigma^{mn})^B{}_{A} [X_m,X_n] \Gamma^{P}\bigg], \label{6dn2susycurrent}\\
\ol{S_+{}^A_P}=&{1\over g_{6}^2} \Tr\bigg[\half \ol{\lambda_-^A} F_{{M}{N}} \Gamma^{P}\Gamma^{MN} +i \ol{\lambda_+^B}(\sigma^{n})^B{}_{A}D_M X_n \Gamma^{P}\Gamma^{M}-\half \ol{\lambda_-^B} (\bar\sigma^{mn})^B{}_{A} [X_m,X_n] \Gamma^{P}\bigg].
}
The $\cN=2$ supersymmetry algebra in six dimensions is computed by dimensional reduction of that of ten dimensional $\cN=1$ algebra calculated in \S\ref{10d}. 
To simplify the situation, we ignore the contributions of fermions. 
SUSY charges are
\be 
Q^A_\pm = \int d^5x S^{0A}_\pm
\ee
and $\Delta\cO=-i[\ol{\epsilon^B_+}Q^B_-+\ol{\epsilon^B_-}Q^B_+,\cO]$, which is justified by the ten dimensional result.
From \eqref{10dlocalsusyalgebra} we calculate the local form of SUSY algebra of $\cN=2$ SYM in six dimensions. 
\beal{ 
\{Q_-^B, \ol{S_-{}^{A}_P}\}
=&  -2i\delta^B\!_A T_{PM}\Gamma^M +J_{PM}\delta^B\!_A \Gamma^M +C^{nm}{}_{P}{}^{QRS} \sigma^{nm}{}^B\!_A \Gamma_{SRQ}, \\
\{Q_+^B, \ol{S_-{}^{A}_P}\}
=& -2i T_{Pm}\sigma^m{}^B\!_A + C^{m}_{PKL}\sigma_m{}^B\!_A \Gamma^{LK} + C^l{}_P{}^{QRST}\sigma_{l}{}^B\!_A \Gamma_{TSRQ}, \\
\{Q_+^B, \ol{S_+{}^{A}_P}\}
=&  -2i\delta^B\!_A T_{PM}\Gamma^M -J_{PM}\delta^B\!_A \Gamma^M - C^{nm}{}_{P}{}^{QRS} \bar\sigma^{nm}{}^B\!_A \Gamma_{SRQ}, \\
\{Q_-^B, \ol{S_-{}^{A}_P}\}
=& 2i  T_{Pm}\bar\sigma^m{}^B\!_A + C^{m}_{PKL}\bar\sigma_m{}^B\!_A \Gamma^{LK} - C^l{}_P{}^{QRST}\bar\sigma_{l}{}^B\!_A \Gamma_{TSRQ},
\label{6dn2localsusyalgebra}
}
where we used the equation of motion of the gauge field 
\be
D_M F^M{}_P - [D_P X^n, X_n]=0, 
\ee
and we set 
\beal{ 
T_{PM}=& {-1\over 4g_6^2} \Tr \big[ g_{MP} (F_{NQ} F^{NQ}+2D_{M}X_m D^{P}X^m -[X_{m},X_n][X^m,X^n])\nn
& +4 (F_{M}{}^N F_{NP}-D_{M}X^m D_P X_{m})  \big],  \\
T_{Pm}=& {i\over g_6^2} \partial_N \Tr[F_P{}^N X_m], \\
C^{nm}{}_{P}{}^{QRS}=&{-i\over 6g_6^2}\varepsilon_P{}^{MNQRS} \partial_M\Tr[X^n D_N X^m], \\
C^{m}_{PKL}=& {-1\over 4g_6^2} \varepsilon_{PRMNKL}\partial^R\Tr \[X^m F^{MN}\], \\
C_l{}_P{}^{QRST}=&{-1\over 72 g_6^2}\varepsilon_P{}^{NQRST}\sigma_{qrml}\partial_N\Tr[[X^q,X^r], X^m].
}
Note that $\sigma_{1234}=1$. 
This leads to supersymmetry algebra of six dimensional $\cN=2$ SYM. 
\beal{ 
\{Q^B_-, \ol{Q^A_-} \} 
=& (-2iP_M+Z_M)\Gamma^M\delta^B_A+Z^{mn}_{QRS}\sigma_{mn}{}^B\!_A\Gamma^{SRQ},\\
\{Q^B_+, \ol{Q^A_-} \} 
=& -2iP_m\sigma^m{}^B\!_A+Z^{m}_{MN}\sigma_{m}{}^B\!_A\Gamma^{NM}+Z^{m}_{QRST}\sigma_{m}{}^B\!_A\Gamma^{TSRQ},\\
\{Q^B_+, \ol{Q^A_+} \} 
=& (-2iP_M-Z_M)\Gamma^M\delta^B_A-Z^{mn}_{QRS}\bar\sigma_{mn}{}^B\!_A\Gamma^{SRQ},\\
\{Q^B_-, \ol{Q^A_+} \} 
=&2iP_m\bar\sigma^m{}^B\!_A+Z^{m}_{MN}\bar\sigma_{m}{}^B\!_A\Gamma^{NM}-Z^{m}_{QRST}\bar\sigma_{m}{}^B\!_A\Gamma^{TSRQ},
}
where 
\beal{ 
P^M =& \int d^5x T^{0M}, \quad P^m = \int d^5x T^{0m}, \\
Z^{mn}_{MNP}=& \int d^5x C^{mn}{}^0{}_{MNP},\quad 
Z^{m}_{KL}= \int d^5x C^{m}{}^0{}_{KL}, \quad
Z^l_{QRST}= \int d^5x C^l{}^0{}_{QRST}.
}

\section{Superalgebra in 5d SYM} 
\label{5d}
In this section we study supersymmetry algebra of SYM in five dimensions. We derive this by dimensional reduction from higher dimensional SUSY algebra studied in the previous sections.

\subsection{Vector multiplet} 
\label{5dVM}
We first consider SYM consisting of one vector multiplet.  
This theory can be obtained by dimensional reduction of six dimensional theory studied in \S\ref{6dVM}. 
We degenerate the fifth direction and the gauge field of this direction becomes a scalar field in the adjoint representation, $ A_5= i \varphi$, where $\varphi$ takes a real value. 
We decompose the gamma matrices in six dimensions into five dimensions ones in the following way. 
\be
\Gamma_\mu = \gamma_\mu \otimes \sigma_1, \quad
\Gamma_5= {\bf 1} \otimes \sigma_2.
\ee
Then the six dimensional chirality matrix and the charge conjugation matrix is computed as 
\beal{
\hat\Gamma =& {\bf 1}\otimes \sigma_3, \quad 
C_{6}=C_5 \otimes i\sigma_2,
}
where $C_{5}=-i\gamma_{03}$. 
The Sp(1)-Majorana Weyl fermion $\lambda^A_{6d}$ in six dimension constrained by \eqref{6dMW} reduces to 
\beal{
\lambda^A_{6d} = 
\left(\begin{array}{c}
\lambda^A \\
\lambda'^A
\end{array}\right), \quad 
\lambda'^A =0, \quad 
\varepsilon^{AB} C_5 \ol{\lambda^B} = \lambda^A,
}
which is simply the Sp(1)-Majorana condition in five dimensions. 

From the six dimensional Lagrangian of $\cN=1$ SYM of a vector multiplet given by \eqref{6dn1Lagrangian}, 
we obtain that of five dimensional one as 
\beal{ 
{\cal L}_V
=&{1\over g_{5}^2}
\Tr \biggl[ {1\over 4}  F_{\mu\nu}  F^{\mu\nu} - {1\over 2} D_\mu \varphi D^\mu \varphi +\half D^A\!_B D^B\!_A +\half \ol{\lambda^A } \slash\!\!\!\! {D} \lambda^A 
-{1\over2} \ol{\lambda^A } [\varphi, \lambda^A]  \biggr]
\label{5dn1Lagrangian}
}
where $g_5$ is a coupling constant of this theory and $\sla \! D=\gamma^\mu D_\mu$. 
The SUSY transformation rule can also be obtained from six dimensional one given by \eqref{6dsusytr}.
\bes{
\Delta  A_\mu =&  \ol{\epsilon^A} \gamma_\mu \lambda^A, \\
\Delta \varphi =& \ol{\epsilon^A} \lambda^A,\\
\Delta \lambda^A =&\half  F_{\mu\nu} \gamma^{\mu\nu}\epsilon^A - \sla D \varphi\epsilon^A+\alpha D^A\!_B \epsilon^B, \\
\Delta D^A\!_B =&\alpha\big( D_\mu \ol{\lambda^B} \gamma^\mu \epsilon^A -[\varphi, \ol{\lambda^B}]\epsilon^A - \half\delta^A_B (D_\mu \ol{\lambda^C} \gamma^\mu \epsilon^C -[\varphi, \ol{\lambda^C}]\epsilon^C) \big),
\label{5dsusytr}
}
where $\epsilon^A$ is a fermionic supersymmetry parameter satisfying $-\varepsilon^{AB} C_5 \ol{\epsilon^B} = \epsilon^A$. $\alpha$ is arbitrary when we consider only a vector multiplet while $\alpha=\sqrt2$ when a hyper multiplet is introduced. 
The supersymmetry current of the Lagrangian  \eqref{5dn1Lagrangian} is computed by dimensional reduction from \eqref{6dsusycurrent}. 
\beal{
\ol{S^A_\rho} =& {1\over g_5^2}\Tr[\ol{\lambda^A}(\half F_{\mu\nu}\gamma^\rho\gamma^{\mu\nu} - D_{\mu}\varphi\gamma^\rho\gamma^{\mu})].
}
The supersymmtery algebra of this theory can be computed in the same way. The supersymmetry variation of the SUSY current is computed as follows.  
\beal{
\Delta\ol{ S^{A}_\rho}
=&  -2 T_{\rho \mu}\ol{\epsilon^A}\gamma^\mu +2i T_{\rho 5}\ol{\epsilon^A} \nn
&+{1\over4g_5^2} (\Tr[ F_{\mu\nu} F_{\sigma\lambda}]\ol{\epsilon^A}\gamma_\rho{}^{\mu\nu\sigma\lambda}+4 \partial_{\sigma } \Tr[ F_{\mu\nu}\varphi]\ol{\epsilon^A}\gamma_\rho{}^{\mu\nu\sigma\lambda}\gamma_\lambda + 2 \partial_{\mu} \Tr[\varphi D_{\sigma} \varphi]\ol{\epsilon^A}\gamma_\rho{}^{\mu \sigma\nu\lambda}\gamma_{\lambda\nu}) \nn
&+{1\over g_5^2} \partial_\mu \Tr[\ol{\lambda^A}\gamma_{\rho}\lambda^B]\ol{\epsilon^B}\gamma^{\mu}-{1\over2g_5^2}  \partial^\nu( \Tr[\ol{\lambda^C}\gamma_{\rho\mu\nu}\lambda^C]\ol{\epsilon^A}\gamma^{\mu}+\Tr[\ol{\lambda^C}(-\gamma_{\rho\nu})\lambda^C]\ol{\epsilon^A})
}
where $\gamma_{01234}=-i$, 
\beal{ 
T_{\mu\rho} =&  {1\over 4g_5^2} 
\Tr \big[ g_{\mu\rho} ( F_{\nu\sigma}  F^{\nu\sigma}-2D_{\nu}\varphi D^{\nu}\varphi) +4 ( F_{\mu}{}^\nu  F_{\nu\rho} +D_{\mu}\varphi D_{\rho}\varphi)  - 2\ol{\lambda^A} \gamma_{(\mu} D_{\rho)}\lambda^A   \big],  \nn
T_{\rho5}=&{i\over g_5^2}\partial_{\nu} \Tr \big[ F_{\rho}{}^\nu \varphi \big]+{1\over 4g_5^2} 
\Tr \big[- i \ol{\lambda^A} \gamma_{\rho} [\varphi,\lambda^A] - i\ol{\lambda^A} D_{\rho}\lambda^A \big].  
}
The supercharge is 
\beal{
Q^A=\int d^4x S^{0A}, 
\label{5dsusycharge}
}
and SUSY transformation is given by $\Delta \cO = [-i\ol{\epsilon^A} Q^A, \cO]$, which is already justified from the higher dimensional result. 
Thus anti-commutation relation of the supercharge and supercurrent is computed as 
\beal{
\{Q^B,\ol{ S^{\rho A}}\}
=& \delta^B_A ( -2i T_{\rho \mu}\gamma^\mu -2 T_{\rho 5} + j_\rho +j'_{\rho}+(j_{\rho\mu}+j'_{\rho\mu}) \gamma^\mu + j_{\rho\nu\lambda} \gamma^{\lambda\nu} ) +j'^{B}_A{}_{\rho\mu}\gamma^{\mu} 
}
where we set 
\beal{
j_\rho=& {i\over4g_5^2} \Tr[ F_{\mu\nu} F_{\sigma\lambda}]\gamma_\rho{}^{\mu\nu\sigma\lambda}, \\
j_{\rho\lambda} =& {i\over g_5^2} \partial_{\sigma } \Tr[ F_{\mu\nu}\varphi]\gamma_\rho{}^{\mu\nu\sigma}{}_\lambda, \\
j_{\rho\nu\lambda} =& {i\over 2g_5^2} \partial_{\mu} \Tr[\varphi D_{\sigma} \varphi]\gamma_\rho{}^{\mu \sigma}{}_{\nu\lambda}, \\
j'^{B}_A{}_{\rho\mu} =& {i\over g_5^2} \partial_\mu \Tr[\ol{\lambda^A}\gamma_{\rho}\lambda^B], \\
j'_{\rho\mu} =& -{i\over2g_5^2} \partial^\nu( \Tr[\ol{\lambda^C}\gamma_{\rho\mu\nu}\lambda^C], \\
j'_{\rho} =& {i\over2g_5^2} \partial^\nu\Tr[\ol{\lambda^C}\gamma_{\rho\nu} \lambda^C].
}
Note that $j_\rho$ is instanton-particle number current. 
Volume integration of both sides gives SUSY algebra of five dimensional SYM as 
\beal{
\{Q^B,\ol{Q^A}\} =& 
\delta^B_A ((-2iP_{\mu} + Z_\mu +Z'{}_\mu) \gamma^\mu -2P_{5} +Z+ Z'+ Z_{\nu\lambda} \gamma^{\lambda\nu}) + Z'^B_A{}_\mu \gamma^\mu
\label{5dsusyalgebra}
}
where  
\beal{
&P^\mu =\int d^4x T^{0\mu}, \quad P^5=\int d^4x T^{05},\\
&Z= \int d^4x j{}^0, \quad 
Z_\lambda=\int d^4x j^0{}_\lambda, \quad 
Z_{\nu\lambda}= \int d^4x j^0{}_{\nu\lambda}, 
\label{5dZ} \\ 
&Z'^B_A{}_\mu= \int d^4x j'^{B}_A{}^0{}_\mu, \quad 
Z'{}_\mu= \int d^4x j'^0{}_\mu, \quad 
Z' = \int d^4x j'^0.
}
$P^5$ is the Kaluza-Klein momentum arising by circle compactification of six dimensional SYM and $Z$ is the instanton-particle charge. These are different but indistinguishable in the superalgebra.

\subsection{Inclusion of a hyper multiplet} 
\label{hyper5d}

We can obtain a theory of a hyper multiplet in five dimensions by dimensional reduction for six dimensional theory studied in \S\ref{6dHM}. 
The chiral fermion in six dimensions denoted by $\psi_{6d}$ reduces to  
\beal{
\psi_{6d}  = 
\left(\begin{array}{c}
\psi'\\
\psi
\end{array}\right), \quad 
\psi'=0.
}
By using this notation we obtain the Lagrangian of a hyper multiplet from \eqref{6dhyperlagrangian}.
\beal{
\cL_{H}{=}& -D_M (q^A)^\dagger D^M q^A - (q^{A})^\dagger \varphi^2 q^{A}   
+ \half \ol{\psi}\slash\!\!\!\! D \psi   +{1\over2} \ol{\psi}\varphi \psi  \nn
&+ \varepsilon^{AB} (q^A)^\dagger  \ol{\lambda^B} \psi  
- \varepsilon_{AB} \ol{\psi} \lambda^A q^B 
+ \sqrt2 (q^A)^\dagger D^A\!_B q^B.
\label{5dhyperlagrangian} 
}
The SUSY transformation is computed from six dimensional one as 
\beal{
\Delta q^A =& \varepsilon^{AB} \ol{\epsilon^B}\psi, \quad  
\Delta \psi = 2 \varepsilon_{BA} (\gamma^\mu  \epsilon^B D_\mu - \epsilon^B \varphi) q^A.
\label{5dsusutr}
}
Five dimensional SUSY current is computed as 
\beal{
\ol{S^A_\rho}
=& {1\over g_5^2}\Tr[\ol{\lambda^A}(\half F_{\mu\nu}\gamma^\rho\gamma^{\mu\nu} - D_{\mu}\varphi\gamma^\rho\gamma^{\mu})] +\varepsilon_{AB} \ol{\psi_-}\gamma_\rho(\gamma^\nu D_\nu q^B-\varphi q^B) \nn
&+(D_\nu q^A)^\dagger \psi^T C_5 \gamma_\rho \gamma^\nu
-(q^A)^\dagger\varphi \psi^T C_5 \gamma_\rho 
-2 (q^A)^\dagger   \ol{\lambda^B} \gamma_\rho  q^B + (q^B)^\dagger \ol{\lambda^A} \gamma_\rho q^B. 
\label{5dsusycurrent2}
}
Supersymmetry variation of supercurrent is computed from \eqref{6ddeltasusycurrent} as 
\beal{
\Delta \ol{ S^A_\rho}
=&-2 T_{\rho \mu}\ol{\epsilon^A}\gamma^\mu +2i T_{\rho 5}\ol{\epsilon^A} \nn
&+{1\over4g_5^2} (\Tr[ F_{\mu\nu} F_{\sigma\lambda}]\ol{\epsilon^A}\gamma_\rho{}^{\mu\nu\sigma\lambda}+4 \partial_{\sigma } \Tr[ F_{\mu\nu}\varphi]\ol{\epsilon^A}\gamma_\rho{}^{\mu\nu\sigma\lambda}\gamma_\lambda +2\partial_{\mu} \Tr[\varphi D_{\sigma} \varphi]\ol{\epsilon^A}\gamma_\rho{}^{\mu \sigma\nu\lambda}\gamma_{\lambda\nu}) \nn
&-2\partial_\mu [(q^A)^\dagger D_\nu q^B-\half \delta^B_A(q^C)^\dagger D_\nu q^C]\ol{\epsilon^B}\gamma_\rho{}^{\mu \nu\sigma\lambda}\gamma_{\lambda\sigma} \nn
&+4\partial_\mu [(q^A)^\dagger \varphi q^B-\half \delta^B_A(q^C)^\dagger \varphi q^C]\ol{\epsilon^B}\gamma_\rho{}^{\mu} 
}
where the stress tensor of the bosonic fields is given by
\beal{ 
T_{\mu\rho} =&  {1\over 4g_5^2} 
\Tr \big[ g_{\mu\rho} ( F_{\nu\sigma}  F^{\nu\sigma}-2D_{\nu}\varphi D^{\nu}\varphi) +4 ( F_{\mu}{}^\nu  F_{\nu\rho}+D_{\mu}\varphi D_{\rho}\varphi)  \big]  \nn
&+2D_{(\mu}(q^A)^\dagger D_{\rho)} q^A -g_{\mu\rho}\partial_\nu((q^A)^\dagger D^\nu q^A), \\
T_{\rho5}=&{i\over g_5^2} \partial_{\nu} \Tr \big[  F_{\rho}{}^\nu \varphi \big] +i D_{\mu}(q^A)^\dagger \varphi q^A - i (q^A)^\dagger\varphi D_{\mu} q^A. 
}
By using $\Delta \cO = [-i\ol{\epsilon^A} Q^A, \cO]$, we obtain
local supersymmetry algebra of five dimensional SYM including a hyper multiplet. 
\beal{ 
\{Q^B,\ol{ S^A_\rho}\}
=& \delta^B_A ( -2 iT_{\rho \mu}\gamma^\mu -2 T_{\rho 5} +  j_\rho +j_{\rho\lambda} \gamma^\lambda + j_{\rho\nu\lambda} \gamma^{\lambda\nu}) +c^B_A{}_{\rho\sigma\lambda} \gamma^{\lambda\sigma}+c^B_A{}_{\rho\nu\sigma\lambda}\gamma^{\lambda\sigma\nu} 
\label{5dlocalsusyhyper}
}
where 
\beal{
c^B_A{}_{\rho\sigma\lambda}=&-2i\partial_\mu [(q^A)^\dagger D_\nu q^B-\half \delta^B_A(q^C)^\dagger D_\nu q^C]\gamma_\rho{}^{\mu \nu}{}_{\sigma\lambda}, \\
c^B_A{}_\rho{}^{\nu\sigma\lambda}=&{2i\over3}\partial_\mu [(q^A)^\dagger \varphi q^B-\half \delta^B_A(q^C)^\dagger \varphi q^C]\gamma_\rho{}^{\mu\nu\sigma\lambda}.
}
Thus supersymmetry algebra in SYM in five dimensions is obtained as 
\beal{
\{Q^B,\ol{Q^A}\} =&
\delta^B_A ( (-2iP_{\mu} + Z_\mu) \gamma^\mu -2 P_{5} +Z+Z_{\nu\lambda} \gamma^{\lambda\nu}) + Y^B_A{}_{\mu\nu\sigma}\gamma^{\sigma\nu\mu} +Y_A^B{}_{\nu\mu} \gamma^{\mu\nu}
\label{5dsusyalgebrahyper}
}
where $Z, Z_\mu, Z_{\nu\lambda}$ are given by \eqref{5dZ}, and 
\beal{
P^\mu =&\int d^4x T^{0\mu}, \quad P^5=\int d^4x T^{05},\quad  
Y^B_A{}^{\nu\mu}=  \int d^4x c^B_A{}^{0\nu\mu}, \quad 
Y_A^B{}^{\nu\sigma\lambda} = \int d^4x c^B_A{}^{0\nu\sigma\lambda}.
}

Note that one can add a real mass parameter $m$ for the hyper multiplet by giving a vev to the adjoint scalar field $\varphi$. Modification of SUSY algebra is done by replacing $\varphi\to m+\varphi$.

\subsection{5d $\cN=2$ superalgebra} 

We study SUSY algebra of maximally supersymmetric Yang-Mills theory in five dimensions in this subsection. 
This was studied in a different notation in \cite{Lambert:2010iw}.  
This theory can be obtained by reducing six dimensional $\cN=2$  SYM to five dimensional one. 
Two Sp(1)-Majorana Weyl fermions in six dimensions denoted by $\lambda^A_\pm$ in \S\ref{6dn2} reduces to  
\beal{
\lambda_+^A = 	
\left(\begin{array}{c}
\lambda_1^A \\
\lambda_1'^A
\end{array}\right), \quad 
\lambda_-^A  = 
\left(\begin{array}{c}
\lambda_2'^A \\
\lambda_2^A
\end{array}\right)
}
with
\beal{
\lambda'^A_1 = \lambda_2'^A =0, \quad 
\varepsilon^{AB} C_5 \ol{\lambda^B_1}^T = \lambda^A_1, \quad 
\varepsilon^{AB} C_5 \ol{\lambda^B_2} = \lambda^A_2.
}
The six dimensional $\cN=2$ Lagrangian given by \eqref{6dn2Lagrangian2} reduces to 
\beal{
\cL^{\cN=2}=& \frac{1}{g_{5}^2} \Tr
\biggl[{1\over 4} F_{\mu\nu} F^{\mu\nu}-\half D_{\mu}\varphi D^{\mu} \varphi -\half D_\mu X_m D^\mu X^m + \half [\varphi,X_m] [\varphi,X^m]+{1\over4}[X_m,X_n][X^m,X^n]\nn
&+{1\over2}\ol{\lambda^A_1}\gamma^\mu D_\mu \lambda^A_1
-{1\over2}\ol{\lambda^A_1} [\varphi,\lambda^A_1]
+{1\over2}\ol{\lambda^A_2} \gamma^\mu D_\mu \lambda^A_2
+{1\over2}\ol{\lambda^A_2} [\varphi,\lambda^A_2] \nn
&+{1\over2} (-i\overline{\lambda^A_{1}} (\bar\sigma_{m})^A\!_B [ X^m, \lambda_{2}^B]
+i\overline{\lambda^A_{2}} (\sigma_{m})^A\!_B [X^m, \lambda_{1}^B])
\biggl].
\label{5dn2Lagrangian} 
}
Five dimensional $\cN=2$ transformation rule is obtained from \eqref{6dn2sym}.
\bes{
\Delta  A_{\mu} =& \ol{\epsilon^A_1} \gamma_{\mu}{\lambda^A_1} + \ol{\epsilon^A_2} \gamma_{\mu}{\lambda^A_2}, \\
\Delta \varphi =&\ol{\epsilon^A_1}{\lambda^A_1} - \ol{\epsilon^A_2} {\lambda^A_2}, \\
\Delta X_{m}
=&i\overline{\epsilon^A_1} (\bar\sigma_{m})^A\!_B \lambda_2^B -i\overline{\epsilon^A_2} (\sigma_{m})^A\!_B  \lambda_1^B, \\
\Delta\lambda^A_1=& \half  F_{\mu\nu}\gamma^{\mu\nu}\epsilon^A_{1} - \sla D \varphi\epsilon^A_{1}
-i\sla D X_m \bar\sigma^m{}^A\!_B\epsilon^B_{2}-i[\varphi, X_m] \bar\sigma^m{}^A\!_B \epsilon^B_{2} -
{1\over 2} [X_m,X_n] \sigma^{mn}{}^A\!_B\epsilon^B_{1}, \\
\Delta\lambda^A_2=& \half  F_{\mu\nu}\gamma^{\mu\nu}\epsilon_{2}^A + \sla D \varphi\epsilon_{2}^A+i \sla D X_m \sigma^m{}^A\!_B\epsilon^B_{1}-i [\varphi, X_m]\sigma^m{}^A\!_B\epsilon^B_{1}- {1\over 2} [X_m,X_n] \bar\sigma^{mn}{}^A\!_B\epsilon^B_{2},
\label{5dn2sym}
}
where $\epsilon_1, \epsilon_2$ are fermionic SUSY parameters satisfying $\varepsilon^{AB} C_5 \ol{\epsilon^B_1}^T =- \epsilon^A_1,\varepsilon^{AB} C_5 \ol{\epsilon^B_2} =- \epsilon^A_2$

One can rewrite the Lagrangian \eqref{5dn2Lagrangian} in a SO(5) symmetric form. 
This can be easily done by dimensional reduction from ten dimensional SYM. 
Degenerating the five directions $x^I$, where $I=5, 6, 7, 8, 9$, we obtain five real scalar fields from the gauge fields of those directions, denoting by $X_I = -i A_I$. 
We decompose the ten dimensional gamma matrices in a way that
\bea
\Gamma_{\mu}=\gamma_{\mu} \otimes \mbf 1 \otimes \sigma_1, \quad 
\Gamma_I=\mbf 1 \otimes \gamma_I \otimes \sigma_2 \,
\eea
where $\gamma_{\mu}\, ({\mu}=0,\cdots, 4)$ are SO(1,4) gamma matrices, $\gamma_I$ are SO(5) gamma matrices.  
Then ten dimensional chirality matrix and 
charge conjugation matrix are computed as 
\beal{
\hat\Gamma_{10} =&\mbf 1\otimes \mbf 1\otimes \sigma_3,\quad
C_{10} =C_5 \otimes \omega \otimes -i\sigma_2 
}
where $\omega=\gamma_{68}$ is an Sp(2) invariant matrix. 
The Majorana-Weyl condition in ten dimensions \eqref{10dMW}  reduces to 
\bea
\lambda = \left(
\begin{array}{c}
\lambda^\mbf A\\
\lambda'^{\mbf A}
\end{array}
\right),
\quad \lambda'^{\mbf A} = 0, \quad
\lambda^\mbf A = \omega^{\mbf A\mbf B}C_5 \overline{\lambda^\mbf B}^T,
\label{sp2Majorana}
\eea
where $\mbf A=1,2,3,4$. In other words, 
a ten dimensional Majorana-Weyl fermion reduces to an Sp(2)-Majorana fermion in five dimensions.
Under this notation the ten dimensional SYM Lagrangian reduces to 
\beal{
\cL^{\cN=2} =& \frac{1}{g_{5}^2} \Tr
\biggl[{1\over 4} F_{{\mu}{\nu}} F^{{\mu}{\nu}}
-\half D_{\mu} X_I D^{\mu} X^I
+{1\over4}[X_I,X_J][X^I,X^J] \nn
& \qquad 
+{1\over2}\ol{\lambda^\mbf A}\gamma^{\mu} D_{\mu} \lambda^\mbf A 
-{1\over2}\ol{\lambda^\mbf A}(\gamma^I)^\mbf A\!_\mbf B [X_I,\lambda^\mbf B] 
\biggl].
\label{5dn2Lagrangian2}
}
This Lagrangian has manifest $SO(5)\simeq Sp(2)$ symmetry. 
To connect the Sp(2)-invariant Lagrangian \eqref{5dn2Lagrangian2} with \eqref{5dn2Lagrangian}, 
one has to decompose the Sp(2)-Majorana fermion into two Sp(1)-Majorana ones, which breaks manifest Sp(2) symmetry.
Realizing the SO(5) gamma matrices $\gamma_I$ by a chiral expression, we can rewrite the Sp(2)-Majorana condition given by \eqref{sp2Majorana} as
\be
\lambda^\mbf A=
\begin{pmatrix} 
\lambda^A_+ \\
\lambda^A_- 
\end{pmatrix}, \quad 
\lambda_\pm^A = \pm \varepsilon^{AB}C_5 \overline{\lambda_\pm^B}^T
\ee
where $A, B=1,2$. 
The Lagrangian \eqref{5dn2Lagrangian2} agrees with that given by \eqref{5dn2Lagrangian} under an identification such that 
\beal{ 
\lambda^A_+=\lambda^A_1, \quad \lambda^A_-=i\lambda^A_2, \quad X^5=\varphi.
\label{idenfitication2}
}

Five dimensional supersymmetry current can be obtained by dimensional reduction from ten dimensional supersymmetry current \eqref{10dsusycurrent}. 
\beal{
\ol{S_\rho^\mbf A}=&{1\over g_{5}^2}\Tr \biggl[\half  F_{\mu\nu}\ol{\lambda^\mbf A} \gamma^{\rho}\gamma^{\mu\nu}- D_\mu X^I \ol{\lambda^\mbf B}(\gamma^I)^\mbf B\!_\mbf A\gamma^{\rho}\gamma^{\mu}-\half [X^I,X^J] \ol{\lambda^\mbf B}\gamma^{\rho}(\gamma^{IJ})^\mbf B\!_\mbf A \biggr]. 
}
The $\cN=2$ supersymmetry algebra in five dimensions is also computed by dimensional reduction of that of ten-dimensional supersymmetry algebra, which keeps manifest $SO(5)$ symmetry. 
We neglect the contributions of fermions for simplicity. 
From \eqref{deltasusycurrent} we calculate 
\beal{
\Delta \ol{S_\rho^{\mbf A}}=&
-2T_{\rho\mu}\ol{\epsilon^\mbf A}\gamma^\mu
-2iT_{\rho I}\ol{\epsilon^\mbf B}(\gamma^I)^\mbf B\!_\mbf A
+{1\over 4g_5^2} \Tr[ F_{\lambda\sigma} F_{\mu\nu}]\varepsilon_\rho{}^{\lambda\sigma\mu\nu}\ol{\epsilon^\mbf A}
-{1\over g_5^2} \gamma_\rho{}^{\lambda\sigma\mu\nu} \partial_\sigma\Tr[X^I F_{\mu\nu}]\ol{\epsilon^\mbf B}\gamma_I\!^\mbf B\!_\mbf A \gamma_\lambda \nn
&+{1\over 2g_5^2}\gamma_\rho{}^{\lambda\sigma\mu\nu}  \partial_\mu\Tr[X^I D_\nu X^J]\ol{\epsilon^\mbf B}\gamma_{IJ}\!^\mbf B\!_\mbf A \gamma_{\lambda\sigma} 
+{1\over 6g_5^2} \partial_\nu\Tr[[X^I,X^J],X^K]\ol{\epsilon^\mbf B}\gamma_\rho\!^\nu\varepsilon^{IJKLM}\gamma_{ML}\!^\mbf B\!_\mbf A
}
where $\varepsilon_{12345}=1$, 
\beal{ 
T_{\mu\rho}
=& {1\over 4g_5^2} \Tr \big[ g_{\mu\rho} ( F_{\nu\sigma}  F^{\nu\sigma}-2D_{\nu}X_I D^{\nu}X^I +[X_{I},X_J][X^I,X^J]) +4 ( F_{\mu}{}^\nu  F_{\nu\rho}+D_{\mu}X^I D_\rho X_{I})   \big],  \nn
T_{\rho I}=&  {i\over g_5^2} \partial_{\nu} \Tr \big[ F_{\rho}{}^\nu X_I\big].
}
The supercharge with Sp(2) index is 
\be 
Q^\mbf A = \int d^4x S^{0\mbf A}, 
\ee
and $\Delta\cO=-i[\ol{\epsilon^\mbf B}Q^\mbf B,\cO]$. 
Then the local supersymmetry algebra of $\cN=2$ SYM is 
\beal{ 
\{Q^\mbf B, \ol{S_\rho^\mbf A} \} 
=& -2iT_{\rho\mu}\delta^\mbf B_\mbf A\gamma^\mu
+2T_{\rho I}(\gamma^I)^\mbf B\!_\mbf A
+j_\rho \delta^\mbf B_\mbf A
+j^I_{\rho\lambda}\gamma_I\!^\mbf B\!_\mbf A \gamma^\lambda \nn
&+j^{IJ}_{\rho\lambda\sigma}\gamma_{IJ}\!^\mbf B\!_\mbf A \gamma^{\lambda\sigma} 
+j^{LM}_{\rho\mu\sigma\lambda} \gamma^{\lambda\sigma\mu}\gamma_{ML}\!^\mbf B\!_\mbf A
}
where 
\beal{ 
j^I_{\rho\lambda}=& -{i\over g_5^2} \gamma_{\rho\lambda}{}^{\sigma\mu\nu} \partial_\sigma\Tr[X^I F_{\mu\nu}], \\
j^{IJ}_{\rho\lambda\sigma}=& {i\over 2g_5^2}\gamma_{\rho\lambda\sigma}{}^{\mu\nu}  \partial_\mu\Tr[X^I D_\nu X^J], \\
j^{LM}_{\rho\mu\sigma\lambda}=&{i\over 36g_5^2} \partial^\nu\Tr[[X_I,X_J],X_K]\gamma_{\rho\nu\mu\sigma\lambda}\varepsilon^{IJKLM}.
} 
Thus supersymmetry algebra is computed as 
\beal{ 
\{Q^\mbf B, \ol{Q^\mbf A} \} 
=& -2iP_\mu \gamma^\mu\delta^\mbf B_\mbf A + 2P_I(\gamma^I)^\mbf B\!_\mbf A +Z \delta^\mbf B_\mbf A
+Z^I_\lambda \gamma_I\!^\mbf B\!_\mbf A \gamma^\lambda \nn
&+Z^{IJ}_{\lambda\sigma} \gamma_{IJ}\!^\mbf B\!_\mbf A \gamma^{\lambda\sigma} +Z^{LM}_{\mu\sigma\lambda} \gamma^{\lambda\sigma\mu}\gamma_{ML}\!^\mbf B\!_\mbf A
\label{5dn2susyalgebra}
}
where we set 
\beal{ 
P^\mu =& \int d^4x T^{0\mu}, \quad
P^I =\int d^4x T^{0I}, \\
Z^I_\lambda=& \int d^4x j^I{}^0{}_{\lambda}, \quad 
Z^{IJ}_{\lambda\sigma}= \int d^4x j^{IJ}{}^0{}_{\lambda\sigma}, \quad 
Z^{LM}_{\mu\sigma\lambda}= \int d^4x j^{LM}{}^0{}_{\mu\sigma\lambda}. 
} 
This result agrees with that in \cite{Lambert:2010iw} up to the conventions, except the quadratic and quartic terms of the scalar fields. 
In \cite{Lambert:2010iw} the quartic term remains as $Z_0^i$, though it always vanishes in our results, which is required for conservation of supercurrent or supercharge. 
Another one is the relative coefficients of the term $X^I D_\nu X^J$ mismatch between those results. 
Note that these mismatches do not affect the analysis done in \cite{Lambert:2010iw}.

\section{Superalgebra in 4d SYM} 
\label{4d} 

In this section we investigate superalgebras in four dimensional SYM by performing dimensional reduction for higher dimensional SYM studied in the previous sections. 
The torus compactification gives four dimensional $\cN=2$ SYM with the kinetic term canonical. 
Thus strictly speaking our study is not so general as to apply to general $\cN=2$ SYM, which admit exact analysis to turn out to have rich structure of SUSY QFT \cite{Seiberg:1994rs,Seiberg:1994aj} and are to be obtained by Riemann surface compactification of six dimensional (2,0) SCFT describing an M-five brane \cite{Witten:1997sc,Gaiotto:2009we}. 
However relying on the Lagrangian description it becomes possible to give an explicit expression for a generic form of $\cN=2$ superalgebra including fermionic contributions. Especially on an instanton or monopole background there exist fermionic zero modes \cite{Jackiw:1975fn,Nohl:1975jg,Callias:1977kg,Atiyah:1978ri,Dorey:2002ik}, which can a priori affect the central charge formula in Coulomb branch \cite{Witten:1978mh}.
Our analysis implies that contribution of such a fermionic zero mode vanishes in the superalgebra. 
In addition our study provides another method to derive extended supersymmetry algebra in four dimensions, which may be simpler than direct computation in canonical formalism especially for the fermionic part \cite{Popescu:2001rf}.

\subsection{$\cN=2$ vector multiplet} 
\label{4dVM}

First we consider $\cN=2$ SYM consisting of an $\cN=2$ vector multiplet, which can be obtained by six dimensional SYM studied in \S\ref{6dVM}. 
We compactify six dimensional theory in $x^4, x^5$ directions. Degeneration of the two tori leads to two real scalar fields from the gauge fields of these direction. 
We combine them to be a complex scalar field so that 
\be 
\phi={ A_4 - i A_5 \over \sqrt2}. 
\label{complexscalar} 
\ee
We decompose six dimensional gamma matrices into four dimensional ones by 
\be
\Gamma_\mu = \gamma_\mu \otimes {\bf 1}, \quad
\Gamma_4 = \hat\gamma\otimes \sigma_1, \quad \Gamma_5= \hat\gamma\otimes \sigma_2
\label{6dgamma}
\ee
where $\hat\gamma = i \gamma_{0123}$ is a chirality matrix in four dimensions. 
Then six dimensional chirality matrix and charge conjugation matrix can be computed as 
\beal{
\hat\Gamma =& \hat\gamma\otimes\sigma_3, \quad 
C_6= C_{4}\hat\gamma \otimes i\sigma_2
}
where $C_{4}=-i\gamma_{03}$. 
Thus six dimensional symplectic-Majorana Weyl fermion constrained by \eqref{6dMW}, which we denote by $\lambda_{6d}$, is decomposed as follows. 
\beal{
\lambda^A_{6d} = 
\left(\begin{array}{c}
\lambda_+^A \\
\lambda_-^A
\end{array}\right), \quad 
\hat\gamma\lambda^A_\pm =\pm \lambda_\pm^A,\quad 
\lambda^A_\pm =\varepsilon^{AB} C_{4} (\ol{\lambda^B_\mp})^T.
\label{4dgaugino}
}
By using the last condition above, we can always write the fermionic part of the theory only in terms of $\lambda^A_+$. 
From \eqref{6dn1Lagrangian} we obtain the Lagrangian of $\cN=2$ SYM as
\beal{ 
{\cal L}_V
=&{1\over g_{4}^2}
\Tr \biggl[\ol{\lambda^A_+} [\slash\!\!\!\! {D},\lambda_+^A] 
+ {1\over 4}  F_{\mu\nu}  F^{\mu\nu} -D_\mu \phi^\dagger D^\mu \phi 
+\half D^A\!_B D^B\!_A - \half  [\phi,\phi^\dagger]^2  \nn
&+{1\over\sqrt2}(-\varepsilon^{AB}  \ol{\lambda^A_+} C_{4}[\phi,\ol{\lambda^B_+}^T] + \varepsilon_{AB}  (\lambda^B_+)^T C_{4}[\phi^\dagger,\lambda^A_+] )
\biggr] 
\label{4dn1Lagrangian} 
}
where $g_4$ is a coupling constant of this theory. 
The $\cN=2$ supersymmetry transformation rule is computed from \eqref{6dsusytr}. 
The supersymmetry parameter denoted by $\epsilon^A_{6d}$ is subject to similar decomposition to gaugino in \eqref{4dgaugino} so that 
\beal{
\epsilon^A_{6d} = 
\left(\begin{array}{c}
\epsilon_+^A \\
\epsilon_-^A
\end{array}\right), \quad 
\hat\gamma\epsilon^A_\pm =\pm \epsilon_\pm^A,\quad 
\epsilon^A_\pm =-\varepsilon^{AB} C_{4} (\ol{\epsilon^B_\mp})^T.
\label{4dsusyparameter}
}
Eliminating $\epsilon^A_-$ by using the last equation of \eqref{4dsusyparameter} we obtain $\cN=2$ SUSY transformation. 
\bes{
\Delta  A_\mu =&  
\ol{\epsilon^A_+} \gamma_\mu \lambda_+^A+ \ol{\lambda^A_+} \gamma_\mu \epsilon_+^A, \\
\Delta \phi =&\sqrt2 \varepsilon_{AB}(\epsilon_+^B)^T C_{4} \lambda_+^A,\\
\Delta \lambda_+^A =&\half  F_{\mu\nu} \gamma^{\mu\nu}\epsilon_+^A + \sqrt2 [\sla D, \phi] \varepsilon^{AB} C_{4} \ol{\epsilon^B_+}^T -[\phi,\phi^\dagger ] \epsilon_+^A +\alpha D^A\!_B \epsilon_+^B, \\
\Delta D^A\!_B =&\alpha(D_\mu \ol{\lambda_+^B} \gamma^\mu \epsilon_+^A-\ol{\epsilon_+^B} \gamma^\mu D_\mu \lambda_+^A-\sqrt2 \varepsilon_{BC} [\phi^\dagger,  (\lambda_+^C)^T]C_{4}\epsilon_+^A +\sqrt2\varepsilon^{AC} [\phi, \ol{\lambda_+^B}]C_{4}\ol{\epsilon^C_+}^T \\
&-\half \delta^A_B(D_\mu \ol{\lambda_+^C} \gamma^\mu \epsilon_+^C-\ol{\epsilon_+^C} \gamma^\mu D_\mu \lambda_+^C-\sqrt2 \varepsilon_{DC} [\phi^\dagger,  (\lambda_+^C)^T]C_{4}\epsilon_+^D +\sqrt2\varepsilon^{DC} [\phi, \ol{\lambda_+^D}]C_{4}\ol{\epsilon^C_+}^T) ), 
\label{4dsusytr}
}
where $\alpha$ is arbitrary when one considers only a vector multiplet though $\alpha=\sqrt2$ when we also consider coupling of a hyper multiplet, which is introduced in the next subsection. 
The supersymmetry current in this theory is computed from \eqref{6dsusycurrent}. The result is 
\bes{ 
\ol{S^A_\rho}=& \Tr[\half F_{\mu\nu}\ol{\lambda_+^A}\gamma^\rho\gamma^{\mu\nu}+\sqrt2\varepsilon_{AB}(\lambda_+^B)^TC_4 D_{\mu}\phi^\dagger\gamma^\rho\gamma^{\mu} -\ol{\lambda_+^A}\gamma^\rho [\phi,\phi^\dagger] ], \\
S^A_\rho =& \Tr[\half F_{\mu\nu}\gamma^{\mu\nu}\gamma^{\rho} \lambda_+^A -\sqrt2 D_{\mu}\phi\gamma^{\mu}\gamma^{\rho}\varepsilon^{AB}C_4(\ol{\lambda_+^B})^T + [\phi,\phi^\dagger] \gamma^{\rho}\lambda_+^B].
\label{4dsusycurrent}
}
From the supercurrent we obtain supercharge in four dimensions
\beal{
Q^A=\int d^3x S^{0A}, \quad 
\label{4dsusycharge}
}
then $\Delta \cO = -i[\ol{\epsilon_+^B} Q^B+\ol{Q^B} \epsilon_+^B, \cO]$. 
Performing dimensional reduction for \eqref{6ddeltasusycurrent} we obtain local form of supersymmetry algebra of four dimensional SYM. 
\beal{
\{Q^B,\ol{ S_\rho^{A}}\}
=& (( -2i T_{\rho \mu}+j_{\rho\mu}+j'_{\rho\mu})\delta^B_A +j^B_A{}_{\rho\mu} )\gamma^{\mu}, \\
\{ \ol{Q^B}^T,\ol{ S_\rho^{A}}\}
=&\varepsilon_{AB}(2iT_{\rho} - j_{\rho} -j'_{\rho}) C_4, \\
\{ \ol{Q^B}^T,({ S_\rho^{A}})^T \}
=& (( -2i T_{\rho \mu}-j_{\rho\mu}+j'_{\rho\mu})\delta^A_B -j^A_B{}_{\rho\mu} )(\gamma^{\mu})^T, \\
\{Q^B,({ S_\rho^{A}})^T\}
=&\varepsilon^{BA} (2i T_{\rho} - j_{\rho} - j'_{\rho})^\dagger C_4,  
}
where 
\beal{ 
T_{\mu\rho}=&  {1\over 4g_4^2} 
\Tr \big[ g_{\mu\rho} ( F_{\nu\sigma}  F^{\nu\sigma}-4 D_{\nu} \phi D^{\nu} \phi^\dagger -2[\phi,\phi^\dagger]^2) +4 ( F_{\mu}{}^\nu  F_{\nu\rho}+2D_{(\mu}\phi D_{\rho)}\phi) \nn
& - 2(\ol{\lambda_+^A} \gamma_{\rho} D_{\mu}\lambda_+^A - D_\mu\ol{\lambda_+^A} \gamma_{\rho} \lambda_+^A )  \big],  \\
T_{\rho} =& - {\sqrt2\over g_4^2}  \partial_{\nu} \Tr [ F_{\rho}{}^\nu\phi^\dagger] +{1\over 2g_4^2}\varepsilon^{AB} \Tr [\ol{\lambda_+^A}C_4 D_{\rho} (\ol{\lambda_+^B})^T] \nn
&+ {\sqrt2 \over 4g_4^2} (\Tr [ \ol{\lambda_+^A} \gamma_{\rho}[\phi^\dagger,\lambda_+^A] - [\phi^\dagger,\ol{\lambda^A_+}]\gamma_\rho\lambda^A_+]), \\
j_{\rho\mu}=&  -{1\over g_4^2}\partial_{\lambda}\Tr[\phi D_{\nu} \phi - \phi^\dagger D_\nu \phi^\dagger]i\sigma_{\rho\mu}\!^{\lambda\nu}, \\
j_{\rho}=&-{\sqrt2 \over g_4^2}\partial_\lambda \Tr[\phi^\dagger F_{\mu\nu}]i\sigma_\rho\!^{\lambda\mu\nu}, \\
j'_{\rho} =&-{i\over g_4^2}\varepsilon^{CD}\partial^\mu\Tr[\ol{\lambda_+^C}\gamma_{\rho\mu}C_4\ol{\lambda_+^D}^T],\\
j^B_A{}_{\rho\mu}=&{2i\over g_4^2}\partial_\mu \Tr[\ol{\lambda_+^A}\gamma_{\rho}\lambda_+^B -\half \delta^B_A \ol{\lambda_+^C}\gamma_{\rho}\lambda_+^C], \\
j'_{\rho\nu}=&-{i\over g_4^2}\sigma_{\rho\nu\mu\sigma}\partial^\mu\Tr[\ol{\lambda_+^C}\gamma^\sigma\lambda_+^C], 
}
with $\sigma_{0123}=-i$. 
Note that in the bosonic terms there exists a brane current which describes one dimensional object (string) as $j_{\rho\mu}$.%
\footnote{Although improvement transformations keeping $\cN=2$ SUSY are not known, those in $\cN=1$ were studied in \cite{Dumitrescu:2011iu}, which suggests that Schwinger terms in a superalgebra can be reabsorbed if it behaves suitably at the boundary. 
Thus since $j_{\rho\mu}, j_\rho, j'_{\rho\mu}$ in our superalgebra are Schwinger terms, these may be reabsorbed into an improvement transformation. However, these are not always removable because, for example, $j_\rho$ measures a background magnetic charge in the Coulomb phase and affects the physical central charge.
We leave a problem to clarify whether other Schwinger terms are physical to future works.    
} 
Performing volume integration for both sides, we obtain supersymmetry algebra in four dimensional $\cN=2$ SYM. 
\beal{
\{Q^B,\ol{ Q^{A}}\}
=& (( -2i P_{\mu}+Z_{\mu}+Z'_{\mu})\delta^B_A +Z^B_A{}_{\mu} )\gamma^{\mu}, \\
\{ \ol{Q^B}^T,\ol{ Q^{A}}\}
=&\varepsilon_{AB}(2i\cP -Z - Z') C_4, \\
\{ \ol{Q^B}^T,({ Q^{A}})^T \}
=& (( -2i P_{\mu}-Z_{\mu}+Z'_{\mu})\delta^A_B - Z^A_B{}_{\mu} )(\gamma^{\mu})^T, \\
\{Q^B,({ Q^{A}})^T\}
=&\varepsilon^{BA} (-2i \cP^\dagger - Z^\dagger - Z'^\dagger) C_4  ,
\label{4dsusyalgebra}
}
where we set 
\beal{
P^\mu =&\int d^3x T^{0\mu}, \quad \cP=\int d^3x T^{0},\quad 
Z= \int d^3x j^0, \quad 
Z_\lambda=\int d^3x j^0{}_\lambda, \quad 
\label{4dZ}\\ 
Z'^B_A{}_\mu =& \int d^3x j'^{B}_A{}^0{}_\mu, \quad 
Z'_\mu= \int d^3x j'^0{}_\mu, \quad 
Z' = \int d^3x j'^0.
}
This result shows that contributions of fermion zero modes to the superalgebra vanishes. This is because a fermionic zero mode (on an instanton background) is essentially given by a shift generated by SUSY transformation \cite{Dorey:2002ik}, $\lambda\sim F_{\mu\nu}\gamma^{\mu\nu}\epsilon_0$ with $\epsilon_0$ a constant spinor, which implies that the zero mode scales as $r^{-2}$ near the boundary $r\sim\infty$.

In particular we obtain the famous formula of central charge as 
\beal{
\{ \ol{Q^B}^T,\ol{ Q^{A}}\}= 2\sqrt2 \varepsilon_{AB} {\cal Z} C_4
}
where 
\be 
{\cal Z}= {1 \over \sqrt2} (i\cP - \half Z - \half Z').
\label{centralcharge}
\ee
For example, let us consider SYM with SU(2) gauge group in the Coulomb branch. 
\be 
\phi = a^* \sigma_3, \quad \lambda^A=0, 
\ee 
where $a$ is a complex number. 
In electrically and magnetically charged background such that 
\be 
n_e={1\over g_4^2} \int d^3x \partial_i (i  f^{i0}), \quad 
n_m= {1\over 4\pi } \int d^3x \partial_i (i\half \varepsilon^{ijk} f_{jk}), \quad 
\label{background}
\ee
where $f_{\mu\nu}=\Tr[F_{\mu\nu}\sigma_3]$, 
one can show that the central charge is computed as 
\be 
{\cal Z}= n_e a+ n_m a_D
\label{centralchargesu2}
\ee
where $a_D=\tau_0 a$ with the holomorphic coupling $\tau_0= {4\pi i \over g_4^2}$.%
\footnote{ The real part of the holomorphic coupling appears once the topological term $F\wedge F$ is introduced.} 
This gives the same formula as in \cite{Seiberg:1994rs}.

\subsection{Inclusion of a hyper multiplet} 
\label{4dHM}  

In this subsection we study four dimensional $\cN=2$ SYM including a hyper multiplet by dimensional reduction. 
We use the same notation for two complex scalar fields $q^A$. 
A six dimensional chiral fermion $\psi_{6d}$ reduces to 
\beal{
\psi_{6d}  = 
\left(\begin{array}{c}
\psi_-  \\
\psi_+ 
\end{array}\right), \quad 
\hat\gamma\psi _\pm =\pm \psi_\pm ,
}
Then the Lagrangian of a hyper multiplet reduces to
\beal{
\cL_{H}
=& -D_\mu (q^A)^\dagger D^\mu q^A
+\half( \ol{\psi_+}  \slash\!\!\!\! {D}\psi_+   + \ol{\psi_-}  \slash\!\!\!\! {D}\psi_-  )  + \sqrt2 (q^A)^\dagger D^A\!_B q^B \nn
&  +\varepsilon^{AB} (q^A)^\dagger \ol{\lambda^B_+} \psi_- + (q^A)^\dagger (\lambda^A_+)^T C_{4} \psi_+ +\varepsilon_{AB}  \ol{\psi_-} \lambda^B_+ q^A - \ol{\psi_+}  C_4(\ol{\lambda^A_+})^T q^A   \nn
& +{1\over\sqrt2}( \ol{\psi_-}  \phi\psi_+  + \ol{\psi_+} \phi^\dagger\psi_- )   - (q^A)^\dagger \{\phi,\phi^\dagger\} q^A.
}
The SUSY transformation boils down to
\beal{
\Delta q^A =& \varepsilon^{AB} \ol{\epsilon^B_+}\psi_- - (\epsilon^A_+)^T C_4\psi_+, \\
\Delta\psi_- =&2  (\gamma^\mu \varepsilon_{BA} \epsilon^B_+ D_\mu q^A + \sqrt2 C_4\ol{\epsilon_+^A} \phi q^A), \\ 
\Delta\psi_+ =&2 (-\gamma^\mu  C_4\ol{\epsilon_+^A}^T D_\mu q^A - \sqrt2  \varepsilon_{BA}\epsilon_+^B \phi^\dagger q^A). 
}
The supercurrent is computed from six dimensional one \eqref{6dsusycurrenthyper}. 
Since the part of a vector multiplet was already computed as  \eqref{4dsusycurrent}, 
we have only to compute the part of a hyper multiplet. 
The result is 
\beal{ 
\ol{S {}^A_\rho{}_{\text{hyp}}}=& \varepsilon_{AB} \ol{\psi_-}\gamma_\rho \gamma^\mu D_\mu q^B
+\varepsilon_{AB}\ol{\psi_+}\gamma_\rho(-\sqrt2\phi^\dagger)q^B +(D_\mu q^A)^\dagger \psi_+^T C_4 \gamma_\rho \gamma^\mu
+(q^A)^\dagger \sqrt2\phi^\dagger \psi_-^T C_4 \gamma_\rho \nn
& -2 (q^A)^\dagger \ol{\lambda_+^B} \gamma_\rho q^B + (q^B)^\dagger \ol{\lambda_+^A} \gamma_\rho q^B, \\
S {}^A_\rho{}_{\text{hyp}}
=& - \gamma^\mu \gamma_\rho C_4 \ol{\psi_+}^T D_\mu q^A +\gamma_\rho C_4 \ol{\psi_-}^T \sqrt2\phi q^A +(D_\mu q^A)^\dagger  \gamma^\mu \gamma_\rho \psi_-
+(q^A)^\dagger \sqrt2\phi \gamma_\rho \psi_+\nn
& -2 (q^B)^\dagger \gamma_\rho \lambda_+^A  q^B + (q^B)^\dagger \gamma_\rho \lambda_+^A q^B.
}
Supercharge in four dimensions is given by 
\eqref{4dsusycharge}. 
From \eqref{6ddeltasusycurrent} we obtain local supersymmetry algebra of four dimensional SYM including the contribution of a hyper multiplet. 
\beal{
\{Q ^B,\ol{ S {}_\rho^{A}}\}
=& (( -2i T_{\rho \mu}+j_{\rho\mu})\delta^B_A +(j^B_A{}_{\rho\mu}+c^B_A{}_{\rho\mu}) )\gamma^{\mu}, \\ 
\{ \ol{Q ^B}^T,\ol{ S {}_\rho^{A}}\}
=&\varepsilon_{AB}(2iT_{\rho} -j_{\rho} ) C_4 -\varepsilon_{CB} c^C_A{}_{\rho\sigma\lambda}C_4\gamma^{\lambda\sigma}, \\
\{ \ol{Q ^B}^T,({ S {}_\rho^{A}})^T \}
=& (( -2i T_{\rho \mu}-j_{\rho\mu})\delta^A_B -(j^A_B{}_{\rho\mu}+c^B_A{}_{\rho\mu}) )(\gamma^{\mu})^T, \\
\{Q ^B,({ S {}_\rho^{A}})^T\}
=&\varepsilon^{BA} (2i T_{\rho} -j_{\rho} )^\dagger C_4 -\varepsilon^{CA}(c^C_B{}_{\rho\sigma\lambda})^\dagger\gamma^{\lambda\sigma}C_4,
}
where 
\beal{ 
T_{\mu\rho}=&  {1\over 4g_4^2} 
\Tr \big[ g_{\mu\rho} ( F_{\nu\sigma}  F^{\nu\sigma}-4 D_{\nu} \phi D^{\nu} \phi^\dagger -2[\phi,\phi^\dagger]^2) +4 ( F_{\mu}{}^\nu  F_{\nu\rho}+2D_{(\mu}\phi D_{\rho)}\phi) \big]\nn  
&+2D_{(\mu}(q^A)^\dagger D_{\rho)} q^A -g_{\mu\rho}\partial_\nu((q^A)^\dagger D^\nu q^A), \\
T_{\rho} =&{-\sqrt2 \over g_4^2}\partial_{\nu} \Tr [ F_{\rho}{}^\nu\phi^\dagger] -\sqrt2 D_{\rho}(q^A)^\dagger \phi^\dagger q^A + \sqrt2(q^A)^\dagger \phi^\dagger D_{\rho} q^A, \\
c^B_A{}^{\rho\lambda}=&4i\partial_\mu [(q^A)^\dagger D_\nu q^B-\half \delta^B_A(q^C)^\dagger D_\nu q^C]\sigma^{\rho\mu \nu\lambda},\\
c^B_A{}^{\rho\sigma\lambda}=&-2i\sqrt2\partial_\mu [(q^A)^\dagger \phi^\dagger q^B-\half \delta^B_A(q^C)^\dagger \phi^\dagger q^C]\sigma^{\rho\mu\sigma\lambda}.
}
This leads to supersymmetry algebra in four dimensional $\cN=2$ SYM. 
\beal{
\{Q ^B,\ol{ Q ^{A}}\}
=& (( -2i P_{\mu}+Z_{\mu})\delta^B_A +(Z^B_A{}_{\mu}+y^B_A{}_{\mu}) )\gamma^{\mu}, \\
\{ \ol{Q ^B}^T,\ol{ Q ^{A}}\}
=&\varepsilon_{AB}(2i\cP -Z ) C_4 -\varepsilon_{CB}y^C_A{}^{\mu\nu}C_4\gamma_{\nu\mu},\\
\{ \ol{Q ^B}^T,({ Q ^{A}})^T \}
=& (( -2i P_{\mu}-Z_{\mu})\delta^A_B - (Z^A_B{}_{\mu}+y^B_A{}_{\mu}) )(\gamma^{\mu})^T,\\
\{Q ^B,({ Q ^{A}})^T\}
=&\varepsilon^{BA} (-2 i\cP^\dagger - Z^\dagger ) C_4  
 - \varepsilon^{CA}(y^C_B{}^{\mu\nu})^\dagger \gamma_{\nu\mu}C_4,
\label{4dsusyalgebrahyper}
}
where we set 
\beal{
y_A^B{}_{\mu} = \int d^3x \; c^B_A{}^0{}_{\mu}, \quad 
y_A^B{}_{\mu\nu} = \int d^3x \; c^B_A{}^0{}_{\mu\nu}.
}
Due to inclusion of a hyper multiplet there appear brane currents $c^B_A{}^{\rho\lambda}, c^B_A{}^{\rho\sigma\lambda}$ in the local form of superalgebra and corresponding brane charges $y^B_A{}^{\lambda}, y_A^B{}_{\mu\nu}$ in the superalgebra.  

One can include a complex mass $m$ of a hyper multiplet in this superalgebra by shifting the adjoint scalar field in the vector multiplet in a way that $\phi \to {m \over \sqrt2} + \phi$. Under this shift $T_\rho \to T_\rho - im J_\rho$, where $J_\rho=i (q^A)^\dagger D_\rho q^A-i D_\rho (q^A)^\dagger q^A$ is the U(1) flavor current. 
Thus the formula of central charge \eqref{centralcharge} is changed as 
\be 
{\cal Z}= {1 \over \sqrt2} (i \cP + m F - \half Z).
\label{centralchargehyper}
\ee
where $F= \int d^3x J^0$ is the U(1) flavor charge.%
\footnote{ In multiple flavor case, this changes as 
$  {\cal Z}= {1 \over \sqrt2} (i \cP + m_i F^i - \half Z),$ where $m_i$ is the mass of $i$th hyper multiplet and $F_i$ is the flavor U(1) charge of the $i$th hyper multiplet. 
}

Let us compute this central charge with SU(2) gauge group in the Coulomb branch. 
The Kaluza-Klein momentum is computed as 
\beal{ 
\cP=& -i\sqrt2 a N_e
}
where we used the equation of motion of the gauge field and set  
\be 
N_e={2\over g_4^2} \int d^3x \partial_i (i  f^{i0}). \quad
\label{electriccharghyper} 
\ee
Remark that the hyper multiplet contributes to the Kaluza-Klein momentum so that the electric charge is twice as great as that in pure $\cN=2$ SYM case with the form of central charge fixed.%
\footnote{ The factor two in \eqref{electriccharghyper} does not depend on the number of hyper multiplets.} 
The Dirac quantization condition requires us to redefine the magnetic charge to be half compared to the pure SYM case. 
\be 
N_m= {1\over 8\pi } \int d^3x \partial_i (i \half \varepsilon^{ijk} f_{jk}). \quad 
\ee
Then $Z$ is computed as 
\beal{ 
Z=- {2\sqrt2 }a_D N_m
}
where $a_D=\tau a$ with $\tau= {8\pi i \over g_4^2}$.
Finally the central charge is obtained as 
\be 
{\cal Z}= N_e a+ N_m a_D + {1 \over \sqrt2} F m 
\label{centralchargesu2hyper}
\ee
which matches the formula given in \cite{Seiberg:1994aj} with the same normalization of the holomorphic coupling $\tau$.
This normalization of the holomorphic coupling is important for the $\cN=2$ SU(2) SYM with four flavors to enjoy SL(2, $\mbf Z$) symmetry \cite{Seiberg:1994aj} as well as to obtain the correct moduli space of $\cN=2$ SU(3) SYM with six flavors as the {\it enlarged} fundamental region in the upper half complex plain, in which the {\it genuine} strong coupling limit exists as $\text{Im}\tau\to0$ \cite{Argyres:2007cn}.

\subsection{$\cN=4$ superalgebra} 

In this final subsection we determine superalgebra in $\cN=4$ SYM by dimensional reduction for ten dimensional SYM. 
By compactifying six directions $x^{3+m}$, where $m=1,2, \cdots, 6,$ the gauge fields of these directions become scalar fields, which we denote by $X_m = -iA_{3+m}$. 
Accordingly we decompose the ten-dimensional gamma matrices as 
\bea
\Gamma_{ \mu}=\gamma_{ \mu} \otimes 1, \quad 
\Gamma_m=\hat\gamma \otimes \gamma_m \,
\eea
where $\gamma_m$ are SO(6) gamma matrices, respectively.  
Since $\cN=4$ SYM has SU(4) R-symmetry, 
it is convenient to rewrite the $SO(6)$ vector representation 
by $SU(4)$ anti-symmetric representation. 
\be
\gamma_{a4}=\half(\gamma_a-i \gamma_{a+3}), \quad
\gamma_{ab} = \varepsilon_{abc} (\gamma_{c4})^\dagger,
\label{so6tosu4}
\ee
where $a,b,c=1,2,3$. These satisfy
\be 
\gamma_{\mbf A\mbf B} =-\gamma_{\mbf B\mbf A}, \quad 
\gamma^{\mbf A\mbf B} = \half \varepsilon^{\mbf A\mbf B\mbf C\mbf D}\gamma_{\mbf C\mbf D} =(\gamma_{\mbf A\mbf B})^\dagger
\ee
where $\mbf A, \mbf B, \mbf C, \mbf D=1,2,3,4.$
We do the same thing for $X^m$. 
$\gamma_{\mbf A\mbf B}$ is explicitly realized as
\bea
\gamma_{\mbf A\mbf B} = \left(
\begin{array}{cc}
0 & - \tilde\rho_{\mbf A\mbf B}\\
\rho_{\mbf A\mbf B}&0\\
\end{array}
\right),
\eea
where 
\be 
(\rho_{\mbf A\mbf B}) ^{\mbf C\mbf D} = \delta_\mbf A ^\mbf C \delta_\mbf B ^\mbf D-\delta_\mbf B^\mbf C \delta_\mbf A ^\mbf D, \quad  (\tilde\rho_{\mbf A\mbf B})_{\mbf C\mbf D}=\varepsilon_{\mbf A\mbf B\mbf C\mbf D}.
\ee
Then the ten-dimensional chirality matrix and 
charge conjugation matrix are computed as 
\beal{
\hat\Gamma_{10} =& \left(
\begin{array}{cc}
\hat\gamma&0\\
0&-\hat\gamma\\
\end{array}
\right), \quad 
C_{10} 
=\left(
\begin{array}{cc}
0 &- C_4\hat\gamma\\
- C_4\hat\gamma &0\\
\end{array}
\right).
}
A ten-dimensional Majorana-Weyl fermion \eqref{10dMW}  
is decomposed as  
\bea
\lambda= \left(
\begin{array}{c}
\lambda_{+\mbf A}\\
\lambda_{-}^\mbf A
\end{array}
\right),
\quad \hat\gamma \lambda_{\pm} = \pm\lambda_{\pm}, \quad
\lambda_-^\mbf A =& C_4 (\overline{\lambda_{+\mbf A}})^T, \quad 
\overline{\lambda_{-}^{\mbf A}} = -\lambda_{+\mbf A}^{T} C_4.
\eea
The SYM Lagrangian in ten dimensions \eqref{10dLagrangian} reduces to
\beal{
\cL^{\cN=4}
=& {1\over g_4^2} \Tr
\biggl[{1\over 4} F_{\mu\nu} F^{\mu\nu}
-\half D_\mu X_{\mbf A\mbf B} D^\mu X^{\mbf A\mbf B}
+{1\over4}[X_{\mbf A\mbf B},X_{\mbf C\mbf D}][X^{\mbf A\mbf B},X^{\mbf C\mbf D}] \nn
&\qquad\qquad+\overline{\lambda_{+\mbf A}} \sla D {\lambda_{+\mbf A}} 
+ \overline{\lambda_{+\mbf C}}[i X_{\mbf C\mbf D},  C_4 (\overline{\lambda_{+\mbf D}})^T ]
-\lambda_{+\mbf A}^{T} C_4 [i X^{\mbf A\mbf B},  \lambda_{+\mbf B}]
\biggl].
\label{4dn4lagrangian}
}
The supersymmetry transformation rule is 
\beal{
\Delta  A_{{\mu}} 
=& \ol{\epsilon_{+\mbf A}} \gamma_{\mu} \lambda_{+\mbf A}
-\epsilon_{+\mbf A}^{T} \gamma_{\mu}^T \ol{\lambda_{+\mbf A}}^T, \nn
\Delta X^{\mbf A\mbf B} =& \varepsilon^{\mbf A\mbf B\mbf C\mbf D} \epsilon_{+\mbf C}^{T} C_4 \lambda_{+\mbf D}
+ \ol{\epsilon_{+\mbf A}}C_4 (\overline{\lambda_{+\mbf B}})^T - \ol{\epsilon_{+\mbf B}}C_4(\overline{\lambda_{+\mbf A}})^T, \nn
\Delta \lambda_{+\mbf A} 
=& \half F_{\mu\nu}\gamma^{\mu\nu} \epsilon_{+\mbf A}
-2i \sla{D} X_{\mbf A\mbf B} C_4 (\overline{\epsilon_{+\mbf B}})^T
-2 [X_{\mbf A\mbf B}, X^{\mbf B\mbf C}] \epsilon_{+\mbf C}, 
\label{susyrs4}
}
where we used 
\beal{
\epsilon= \left(
\begin{array}{c}
\epsilon_{+\mbf A}\\
\epsilon_{-}^\mbf A
\end{array}
\right),
\quad \hat\gamma \epsilon_{\pm} = \pm\epsilon_{\pm}, \quad
\epsilon_-^\mbf A =&- C_4 (\overline{\epsilon_{+\mbf A}})^T, \quad 
\overline{\epsilon_{-}^{\mbf A}} = \epsilon_{+\mbf A}^{T} C_4.
}

Let us perform dimensional reduction for SUSY current. 
The result is 
\beal{ 
\ol{S^\rho_{-\mbf A}}=& {1\over g_{4}^2}\Tr \biggl[\half  F_{\mu\nu}\ol{\lambda_{+\mbf A}} \gamma^{\rho}\gamma^{\mu\nu}
-2i D_\mu X^{\mbf B\mbf A} \ol{\lambda^\mbf B_{-}} \gamma^{\rho}\gamma^{\mu}
+ 2 [X_{\mbf B\mbf C},X^{\mbf C\mbf A}] \ol{\lambda_{+\mbf B}}\gamma^{\rho} \biggr], \\
\ol{S^{\rho \mbf A}_{+}}=& {1\over g_{4}^2}\Tr \biggl[\half  F_{\mu\nu}\ol{\lambda_{-}^\mbf A} \gamma^{\rho}\gamma^{\mu\nu}
-2i D_\mu X_{\mbf B\mbf A} \ol{\lambda_{+\mbf B}}\gamma^{\rho}\gamma^{\mu}
+2 [X^{\mbf B\mbf C},X_{\mbf C\mbf A}] \ol{\lambda_{-}^\mbf B}\gamma^{\rho} \biggr].
}
$S^\rho_{-\mbf A}, S^{\rho \mbf A}_{+}$ are determined so as to satisfy $\ol{\epsilon_{-}^\mbf B}S_{+}^{\rho \mbf B} = \ol{S_{-\mbf B}^{\rho}}\epsilon_{+\mbf B}, \ol{\epsilon_{+ \mbf B}}S_{-\mbf B}^{\rho} = \ol{S_{+}^{\rho\mbf B}}\epsilon_{-}^\mbf B$. 
Then the supercharge with SU(4) index is 
\be 
Q_{-\mbf A} = \int d^3x S^{0}_{-\mbf A},\quad  
Q_{+}^\mbf A = \int d^3x S^{0\mbf A}_{+},\quad  
\ee
and $\Delta\cO=-i[\ol{\epsilon_{+\mbf B}}Q_{-\mbf B}+\ol{\epsilon_{-}^\mbf B}Q_{+}^\mbf B,\cO]$.
The $\cN=4$ superalgebra in four dimensions can be computed by dimensional reduction from ten dimension as done in five dimensions. We neglect the contribution of fermions, which is given by total derivative terms and thus vanishes as discussed in the pure $\cN=2$ SYM. 
The local version of supersymmetry algebra of $\cN=4$ SYM is 
\beal{ 
\{Q_{-\mbf B}, \ol{S_\rho{}_{-\mbf A}} \} 
=& -2iT_{\rho\mu}\delta^\mbf A_\mbf B\gamma^\mu
+j^\mbf A_\mbf B{}_{\rho\mu}\gamma^\mu, \\
\{Q_{+}^\mbf B, \ol{S_\rho{}_{-\mbf A}} \} 
=&-2iT_{\rho }^{\mbf B \mbf A}
+j_\rho ^{\mbf B \mbf A}
+j^{\mbf B\mbf A}_{\rho\sigma\lambda} \gamma^{\lambda\sigma}, \\
\{Q_{+}^\mbf B, \ol{S_\rho{}^\mbf A_{+}} \} 
=& -2iT_{\rho\mu}\delta^\mbf B_\mbf A\gamma^\mu
-j^\mbf B_\mbf A{}_{\rho\mu}\gamma^\mu, \\
\{Q_{-\mbf B}, \ol{S_\rho{}^\mbf A_{+}} \} 
=&-2iT_{\rho }{}_{\mbf B \mbf A}-j_\rho {}_{\mbf B \mbf A}
-j_{\mbf B\mbf A}{}_{\rho\sigma\lambda} \gamma^{\lambda\sigma},
}
where 
\beal{ 
T_{\mu\rho}
=& {1\over 4g_4^2} \Tr \big[ g_{\mu\rho} ( F_{\nu\sigma}  F^{\nu\sigma}-2D_{\nu}X_{\mbf A\mbf B} D^{\nu}X^{\mbf A\mbf B} +[X_{\mbf A\mbf B},X_{\mbf C\mbf D}][X^{\mbf A\mbf B},X^{\mbf C\mbf D}]) \nn
& +4 ( F_{\mu}{}^\nu  F_{\nu\rho}+D_{\mu}X^{\mbf A\mbf B} D_\rho X_{\mbf A\mbf B})   \big],  \\
T_{\rho \mbf A\mbf B}=&  {2i\over g_4^2} \partial_{\nu} \Tr \big[ F_{\rho}{}^\nu X_{\mbf A\mbf B}\big], \\
j^\mbf A_\mbf B{}_{\rho\mu}=& -{4i\over g_4^2} \sigma_{\rho\sigma\nu\mu} \partial^\sigma\Tr[X_{\mbf B\mbf C} D^\nu X^{\mbf C\mbf A}], \\
j^{\mbf B\mbf A}_{\rho}=& -{2\over g_4^2}\sigma_\rho{}^{\sigma\mu\nu}  \partial_\sigma\Tr[X^{\mbf B\mbf A}  F_{\mu\nu}], \\
j^{\mbf B\mbf A}_{\rho\sigma\lambda}=&-{4\over 3g_4^2} \sigma_{\rho\nu\sigma\lambda}\partial^\nu\Tr[[X^{\mbf B\mbf C},X_{\mbf C\mbf D}],X^{\mbf D\mbf A}].
} 
Volume integration of both sides leads to supersymmetry algebra in $\cN=4$ SYM. 
\beal{ 
\{Q_{-\mbf B}, \ol{Q_{- \mbf A}} \} 
=& -2iP_{\mu}\delta^\mbf A_\mbf B\gamma^\mu
+Z^\mbf A_\mbf B{}_{\mu}\gamma^\mu, \\
\{Q_{+}^\mbf B, \ol{Q_{- \mbf A}} \} 
=&-2iP^{\mbf B \mbf A}
+Z^{\mbf B \mbf A}
+Z^{\mbf B\mbf A}_{\sigma\lambda} \gamma^{\lambda\sigma}, \\
\{Q_{+}^\mbf B, \ol{Q^\mbf A_{+}} \} 
=& -2iP_{\mu}\delta^\mbf B_\mbf A\gamma^\mu
-Z^\mbf B_\mbf A{}_{\mu}\gamma^\mu, \\
\{Q_{-\mbf B}, \ol{Q^\mbf A_{+}} \} 
=&-2iP_{\mbf B \mbf A}-Z_{\mbf B \mbf A}
-Z_{\mbf B\mbf A}{}_{\sigma\lambda} \gamma^{\lambda\sigma},
\label{4dn4susyalgebra}
}
where we set 
\beal{ 
P^\mu =& \int d^3x T^{0\mu}, \quad
P^{\mbf A\mbf B} =\int d^3x T^{0\mbf A\mbf B}, \\
Z^\mbf A_\mbf B{}_\mu=&\int d^3x j^\mbf A_\mbf B{}^0{}_\mu, \quad 
Z^{\mbf B \mbf A}= \int d^3x j^{\mbf B\mbf A}{}^0, \quad 
Z^{\mbf A\mbf B}_{\lambda\sigma}= \int d^3x j^{\mbf A\mbf B}{}^0{}_{\lambda\sigma}. 
} 

As an example, let us consider the case of SU(2) gauge group in the Coulomb branch. 
\be 
X_{12} = {a\over \sqrt2} \sigma_3, ~~ X_{13}=X_{14} =0.
\ee
Then \eqref{4dn4susyalgebra} is computed as
\be 
\{Q_{-1}, \ol{Q^2_{+}} \} = \{Q_{-3}, \ol{Q^4_{+}} \} =   2\sqrt2 {\cal Z} 
\ee
where 
\be 
{\cal Z}= n_e a+ n_m a_D
\label{centralchargesu2n4}
\ee
with $n_e, n_m$ given by \eqref{background}, $a_D=\tau_0 a$, $\tau_0= {4\pi i \over g_4^2}$. 
The formula of the central charge with normalization of the holomorphic coupling in $\cN=4$ SYM is the same as pure $\cN=2$ SYM, which is again consistent with the result in \cite{Seiberg:1994aj}.

\section{Discussion} 
\label{discussion}

We have determined supersymmetry algebra of SYM of a vector multiplet in six dimensions including the contribution of fermions, which is given by boundary terms. We have extended this calculation to the case including a hyper multiplet. 
For SUSY algebra of six dimensional maximally SYM we have carried out dimensional reduction for that in ten dimensions. 
From six dimensional results we have performed dimensional reduction to determine SUSY algebras of five and four dimensional SYM. 
From six to five the Kaluza-Klein momentum arising from torus compactification is different from the instanton-particle charge though they are indistinguishable in the superalgebra.  
And the Kaluza-Klein momentum corresponds to the electric charge part in the famous formula of central charge. 
We have derived the whole extended supersymmetry algebra as well as the holomorphic coupling constant introduced in \cite{Seiberg:1994aj} against the four dimensional $\cN=2$ SYM including fundamental hyper multiplets and $\cN=4$ SYM. 

Since we started from SYM in six dimensions with the canonical kinetic term in this paper, the theory obtained by dimensional reduction inherited this property. 
Computing SUSY algebra of general SYM with the non-canonical kinetic term is left to future work, though the general structure of the algebra will remain unchanged. 
Especially in five and four dimensions a general Lagrangian contains topological terms such as Chern-Simons term and $F\wedge F$, respectively,
which has an extra effect on physics of the theory \cite{Witten:1979ey}. 

It should be possible to determine BPS states in maximally SYM in six dimensions in Higgs branch. 
In terms of brane picture, maximally SYM in six dimensions is realized on D-five branes, and Higgs branch corresponds to separation thereof. 
Then BPS states on this branch will correspond to supersymmetric brane configuration realized on the separated D-five branes set up. 
It would be interesting to clarify relations between those BPS states in six dimensional SYM and those in five dimensional maximally SYM in broken phase, which has close relationship with the (2,0) theory describing M-five branes \cite{Lambert:2010iw}. 

We hope to come back to these problems in the future.

\section*{Acknowledgments} 

The author was supported by the Israeli Science Foundation under grant 352/13 and 504/13.

\appendix 

\section{A formula of gamma matrix} 
\label{gamma}

In this appendix, we derive a formula of gamma matrix given in Appendix of \cite{Kugo:2000hn}.
We use notation such that
\be
\Gamma^{\mu_0\cdots\mu_n}=\Gamma^{[\mu_0}\cdots\Gamma^{\mu_n]}={1\over (n+1)!} \sum_{\sigma\in \mf S_{n+1}}(-)^\sigma\Gamma^{\mu_{\sigma(0)}}\cdots \Gamma^{\mu_{\sigma(n)}}
\ee
where $\mf S_{N}$ is the set of permutation of $N$ elements. 

Denoting $C_{n,m},D_{n,m}$ by
\beal{
\Gamma^{M_1\cdots M_m}\Gamma^{N_1\cdots N_n}\Gamma_{M_m\cdots M_1}=&C_{n,m}\Gamma^{N_1\cdots N_n},\\
(-)^{m-1}\Gamma^{M_1\cdots M_{m-1}[N_1}\Gamma^{N_2\cdots N_n]}\Gamma_{M_{m-1}\cdots M_1}=&D_{n,m}\Gamma^{N_1\cdots N_n},
}
we can relate $C_{n,m}, D_{n,m}$ by
\beal{
C_{n,m}=&(-)^m C_{n-1,m}+2m(-)^{m+n}D_{n-1,m-1},\\
D_{n,m}=&\half(C_{n,m}+(-)^mC_{n+1,m}).
}
From these we find 
\beal{
C_{n,m}=(-)^mC_{n-1,m}+m(-)^{m+n}C_{n-1,m-1}+m(-)^{n+1}C_{n,m-1}.
\label{Cnm}
}
One can easily check that 
\beal{
&C_{n,0}=1, \quad C_{0,m}=D(D-1)\cdots(D-(m-1)), 
\label{initialCnm}
}
where $D$ is an arbitrary dimension.
By using \eqref{Cnm} and \eqref{initialCnm}
one can determine $C_{n,m}$ (and thus $D_{n,m}$) inductively. 
As examples, we determine $C_{n,m}$ when $D=6,10$.
\begin{table}[htbp]
\begin{center}
\begin{tabular}{|c|c|c|c|c|c|}
\hline
$m=3$ & 120 & 0 &$-24$ & 0 \\ 
\hline
$m=2$ & 30 & 10 & $-2$ & $-6$\\ 
\hline
$m=1$ & 6 & $-4$  & 2 & 0\\ 
\hline
$m=0$ & 1 & 1 & 1 & 1\\
\hline
$C_{n,m}$ & $n=0$ & $n=1$ & $n=2$ & $n=3$\\
\hline
\end{tabular} 
\caption{$D=6$.}
\end{center}
\end{table} 
\begin{table}[htbp]
\begin{center}
\begin{tabular}{|c|c|c|c|c|c|c|c|}
\hline
$m=5$ &30240 & 0 &$-3360$ & 0 & 1440 & 0 \\ 
\hline
$m=4$ & 5040 &1008 &$-336$ & $-336$ & 48 &240 \\ 
\hline
$m=3$ & 720 & $-288$ &$48$ & $48$ & $-48$ & 0 \\ 
\hline
$m=2$ & 90 & 54 & $26$ & $6$ & $-6$ & $-10$ \\ 
\hline
$m=1$ & 10 & $-8$ &6 & $-4$ & 2 & 0\\ 
\hline
$m=0$ & 1 & 1 & 1 & 1 & 1 & 1 \\
\hline
$C_{n,m}$ & $n=0$ & $n=1$ & $n=2$ & $n=3$& $n=4$& $n=5$\\
\hline
\end{tabular} 
\caption{$D=10$.}
\end{center}
\end{table} 
The result of $D=6$ matches that given in \cite{Kugo:2000hn}.

\section{Convention in six dimensions}
\label{convention}
In this appendix we collect convention in six dimensions used in this paper. 
We realize SO(1,5) matrices satisfying $\{\Gamma_M, \Gamma_N\}=2g_{MN}$, where $(g_{MN})=\text{diag}(-1, 1, \cdots, 1)$, in two ways. 
One is 
\be
\Gamma_\mu = \gamma_\mu \otimes {\sigma_1}, \quad \Gamma_5= {\bf 1}\otimes \sigma_2
\label{6dgamma1}
\ee
where $\gamma^\mu$ $(\mu=0,1,2,3)$ are SO(1,3) matrices realized as
\beal{
\gamma_\mu= 
\begin{pmatrix}
0 & \bar\sigma'_\mu \\
\sigma'_\mu & 0
\end{pmatrix}, \quad 
} 
where $\bar\sigma'_0=\sigma'_0=i\sigma_2$, $\bar\sigma'_1=\sigma'_1=\sigma_1$, $\bar\sigma'_2=\sigma'_2=\sigma_3$, $\bar\sigma'_3=-\sigma'_3=i$ with $\sigma_i$ Pauli matrices satisfying $\sigma_i\sigma_j=\delta_{ij}+i\varepsilon_{ijk}\sigma_k$.  
This realization is useful when we consider dimensional reduction from six dimensions to five ones. 
The other is
\be
\Gamma_\mu = \gamma_\mu \otimes {\bf 1}, \quad
\Gamma_4 = \hat\gamma\otimes \sigma_1, \quad \Gamma_5= \hat\gamma\otimes \sigma_2
\label{6dgamma2}
\ee
where $\hat\gamma = i \gamma_{0123}$ is a chirality matrix in four dimensions. 
This is convenient when we do dimensional reduction from six to four.
In both cases, we define the charge conjugation matrix as 
$
C_6=\Gamma_{035},
$
which satisfies 
\bea
C_6^2 =1, \quad C_6^* =  C_6, \quad
C_6^T =C_6,\quad C_6 \Gamma^M = - (\Gamma^M)^T C_6. 
\eea

In Lorentzian six dimensions 
there exists a symplectic majorana Weyl spinor. 
By denoting Sp(1)-Majorana fermion by $\lambda^A$ it satisfies
\be
\lambda^A=\varepsilon^{AB} C_6 \ol{\lambda^B}^T.
\ee
Two symplectic Majorana fermions $\psi^A,\chi^A$ satisfy  
\be
\ol{\psi^A} \gamma_{\mu_1}\cdots\gamma_{\gamma_k}\chi^B
=(-)^{k+1}(\overline{\chi^{A}}  \gamma_{\mu_k}\cdots\gamma_{\gamma_1}\psi^B -\delta^B_A \overline{\chi^D} \gamma_{\mu_k}\cdots\gamma_{\gamma_1}\psi^D).
\ee
Especially taking trace in terms of Sp(1) index gives 
\be
\ol{\psi^A} \gamma_{\mu_1}\cdots\gamma_{\gamma_k}\chi^A
=(-)^{k}\overline{\chi^A} \gamma_{\mu_k}\cdots\gamma_{\gamma_1} \psi^A.
\ee

\bibliographystyle{utphys}
\bibliography{6dsusyv2}
\end{document}